\pdfoutput=1

\documentclass[prb,amsmath,amssymb,amsbsy,twocolumn,superscriptaddress,showpacs]{revtex4-1}

\usepackage[svgnames]{xcolor}
\definecolor{darkblue}{RGB}{0 60 120}
\definecolor{eggplant}{RGB}{190 10 150}
\definecolor{darkgray}{RGB}{70 70 70}
\usepackage{amsfonts}
\usepackage{hhline}
\usepackage{amsmath}
\usepackage{amssymb}
\usepackage{bm}
\usepackage{graphicx}
\usepackage[breaklinks, colorlinks = true,linkcolor=eggplant,urlcolor=darkblue,citecolor=eggplant]{hyperref}
\hypersetup{
    colorlinks,
    citecolor=black,
    filecolor=black,
    linkcolor=black,
    urlcolor=black
}
\usepackage{mathrsfs}
\usepackage[lofdepth,lotdepth,caption=false]{subfig}
\usepackage{varwidth}
\usepackage{wrapfig}
\usepackage{times}
\usepackage{bm}
\usepackage[percent]{overpic}
\usepackage{longtable}
\usepackage{multirow}
\usepackage{color}

\newcommand{\nairo}{Na${}_{2}$IrO${}_{3}$}
\newcommand{\aliiro}{$\alpha$-Li${}_{2}$IrO${}_{3}$}
\newcommand{\bliiro}{$\beta$-Li${}_{2}$IrO${}_{3}$}
\newcommand{\gliiro}{$\gamma$-Li${}_{2}$IrO${}_{3}$}
\newcommand{\liiro}{Li${}_{2}$IrO${}_{3}$}
\newcommand{\airo}{$A_{2}$IrO${}_{3}$}

\newcommand{\ttg}{$t_{2g}$}
\newcommand{\soc}{SOC}

\newcommand{\ket}[1]{\left|#1\right\rangle}
\newcommand{\ir}{$\text{Ir}^{4+}$}
\newcommand{\jhalf}{$j_{\text{eff}}=1/2$}

\newcommand{\jthreehalf}{$j_{\text{eff}}=3/2$}

\newcommand{\harzero}{$\mathcal{H}\text{--}0$}
\newcommand{\har}{$\mathcal{H}\text{--}1$}

\graphicspath{{figs/}}

\begin{document}

\title{Recent progress on correlated electron systems with strong spin-orbit coupling}


\author{Robert Schaffer}
\author{Eric Kin-Ho Lee}
\affiliation{Department of Physics and Center for Quantum Materials,
University of Toronto, Toronto, Ontario M5S 1A7, Canada.}
\author{Bohm-Jung Yang}
\affiliation{Center for Correlated Electron Systems, Institute for Basic Science (IBS), Seoul 151-747, Korea} \affiliation{Department of Physics and Astronomy, Seoul National University (SNU), Seoul 151-747, Korea}
\author{Yong Baek Kim}
\affiliation{Department of Physics and Center for Quantum Materials,
University of Toronto, Toronto, Ontario M5S 1A7, Canada.}
\affiliation{Canadian Institute for Advanced Research, Quantum Materials Program,
Toronto, Ontario MSG 1Z8, Canada}
\affiliation{School of Physics, Korea Institute for Advanced Study, Seoul 130-722, Korea.}


\begin{abstract}
Emergence of novel quantum ground states in correlated electron systems with strong spin-orbit coupling has been a recent subject of intensive studies. While it has been realized that spin-orbit coupling can provide non-trivial band topology in weakly interacting electron systems, as in topological insulators and semi-metals, the role of electron-electron interaction in strongly spin-orbit coupled systems has not been fully understood. The availability of new materials with significant electron correlation and strong spin-orbit coupling now makes such investigations possible. Many of these materials contain 5d or 4d transition metal elements; the prominent examples are iridium oxides or iridates. In this review, we succinctly discuss recent theoretical and experimental progress on this subject. After providing a brief overview, we focus on pyrochlore iridates and three-dimensional honeycomb iridates. In pyrochlore iridates, we discuss the quantum criticality of the bulk and surface states, and the relevance of the surface/boundary states in a number of topological and magnetic ground states, both in the bulk and thin film configurations. Experimental signatures of these boundary and bulk states are discussed. Domain wall formation and strongly-direction-dependent magneto-transport are also discussed. Regarding the three-dimensional honeycomb iridates, we consider possible quantum spin liquid phases and unusual magnetic orders in theoretical models with strongly bond-dependent interactions. These theoretical ideas and results are discussed in light of recent resonant X-ray scattering experiments on three-dimensional honeycomb iridates. We also contrast these results with the situation in two-dimensional honeycomb iridates. We conclude with the outlook on other related systems.
\end{abstract}
\date{\today}
\maketitle

\tableofcontents

\vspace{5mm}

\section{Introduction}

Recent discoveries of topological insulators and semi-metals have taught us the importance of the spin-orbit coupling in the emergence of non-trivial quantum ground states~\cite{TI_Moore, TI_Kane,TI_Zhang,TI_Ando}. While most of these materials are weakly interacting electron systems, and hence can be understood in terms of an independent electron picture, the effects of electron-electron interactions on such quantum ground states and possible materials where such effects can be seen are currently subjects of great interest~\cite{interactingTI_Raghu,interactingTI_actinide,interactingTI_fractional}. On the other hand, one of the traditional playgrounds for strong correlation effects is a 3d transition metal oxides system, where the narrow bands of 3d orbitals make the electron-electron interaction greatly pronounced~\cite{MIT_Tokura}. This leads to a variety of Mott insulators, broken symmetry orders, and metal-insulator transitions. It has only recently been realized that the strong spin-orbit coupling in 5d transition metal oxides makes this class of materials particularly novel platforms for cooperative effects of electron correlation and strong spin-orbit coupling~\cite{kim2008novel}. This is in contrast to a naive picture of weak correlation effects in 5d transition metal oxides via the wide bands expected from spatially extended 5d orbitals. Numerous new 5d transition metal oxides have been synthesized and studied in the last 5-10 years. This raises the hope that these systems can be used to understand the emergence of novel quantum ground states that are possible only when both the electron interaction and spin-orbit coupling are present.

In this review article, we will outline recent and ongoing progress in this interesting and rapidly developing field. Two recent review articles have given a pedagodgical introduction to the physics of correlated electron systems with strong spin-orbit coupling~\cite{William_review,rau2015spin}; we will not repeat this here, but rather focus on recent advances which have been made,
but not covered or only partially-covered in those articles. However, it is useful to introduce the basic physics which gives rise to these phenomena, which is briefly explained below.

The key effect of dominant SOC in a material is the entanglement between spin and orbital degrees of freedom in the underlying electronic wavefunctions. In strongly correlated materials where an atomic description is typically employed as a starting point, such entanglement necessitates the construction of spin-orbit entangled orbitals. Some of the first classes of materials that clearly demonstrated the aforementioned physics are iridium oxides, or \textit{iridates}~\cite{kim2008novel,structure_Maeno,MIT}. Various energy scales, including on-site Coulomb repulsion, crystal field splitting, and spin-orbit coupling are often comparable and competing with each other in these systems~\cite{Pesin,Yang_pyrochlore}. The interplay of multiple comparable energy scales is the central ingredient that generates the variety of physical phenomena observed in iridate materials. The common motif in these materials is the IrO$_6$ octahedra: each iridium \ir{} ion in its $5d^5$ valence state is surrounded by an octahedral cage of oxygen O$^{2-}$ anions.\cite{} Since the $5d$ orbitals of iridium are spatially extended, one of the important energy scales acting on the $5d$ orbitals arises from crystal field effects of this local environment.  The strong spin-orbit coupling and pronounced crystal field effect allow us to use a simplified description of the localized picture of the iridium ions, which was first elucidated by Kim \textit{et al.}\cite{kim2008novel}. Here we briefly recount this localized picture.

In octahedral crystal field environments, the ten-fold degenerate $d$ orbitals (spin degeneracy included) are split into two sets of multiplets.  The lower energy manifold is six-fold degenerate and the orbital components of the wavefunctions transform like the basis functions $yz$, $xz$, and $xy$.  These orbital components behave like a pseudovector under symmetry operations and are conventionally denoted as the \ttg{} orbitals.  The higher energy manifold is four-fold degenerate and the orbitals transform as a doublet; they are denoted as the $e_g$ doublet.  Since the Ir$^{4+}$ valence state implies that there are five electrons in the $d$ shell, in the limit that the octahedral crystal splitting is large, one ends up with a partially-filled $t_{2g}$ orbitals and an $e_g$ manifold which can be projected out. This leaves five electrons or, equivalently, a single-hole in the \ttg{} manifold. The projection of the angular momentum of the $d$ electrons into the \ttg{} manifold leads to an effective angular momentum one operator with an extra minus sign, ${\vec L}_{\rm eff} = -{\vec L}_{\ell=1}$~\cite{Abragam,Hosub,Pesin}.
The resulting spin-orbit coupling, $-\lambda {\vec L}_{\ell=1} \cdot {\vec S}$ splits the \ttg{} multiplet into an effective \jhalf{} doublet and an effective \jthreehalf{} quartet.  The \jthreehalf{} quartet is lower in energy and separated from the \jhalf{} doublet by a gap of $3\lambda/2$.
As explained later in the review, these pseudo-spin degrees of freedom are used as building blocks of electronic structure and effective interactions in many iridate materials.

In the bulk of this review, we will focus our attention on two classes of materials in which SOC leads to new and interesting physics. First, we will consider the pyrochlore iridate materials with the chemical formula R$_2$Ir$_2$O$_7$, where R is a rare-earth element. Substantial theoretical and experimental progress has been made in understanding the phases which appear in these systems. Experimentally, the all-in/all-out (AIAO) magnetic order which appears in Eu$_2$Ir$_2$O$_7$ and Nd$_2$Ir$_2$O$_7$ has been revealed by elastic and inelastic resonant X-ray scattering experiments~\cite{Nakatsuji_xray,xray2} as well as magneto-transport, magnetization measurements~\cite{Ueda_Bfield,Ueda_optical,Ueda_DW,tian2015field}, and neutron scattering~\cite{Tomiyasu_neutron}. There also exists experimental evidence for magnetic domain walls associated with the AIAO order in Nd$_2$Ir$_2$O$_7$~\cite{Ueda_optical,Ueda_DW,DW_Shen}. In Pr$_2$Ir$_2$O$_7$, the quadratic band touching dispersion has been confirmed in ARPES measurements~\cite{QBC_Arpes}, which is crucial for the explanation of the emergence of Weyl semi-metal when time-reversal symmetry is broken~\cite{Weyl_iridate}. On the theoretical front, in the bulk, the effects of the Coulomb interaction in the semi-metal phases and near various quantum critical points have been studied~\cite{Moon,Savary,Yang_2014}. The nature of the surface states, and in particular the Fermi arc connecting the Weyl points has been examined~\cite{Weyl_iridate}, and the possibility of finding a large anomalous Hall effect in thin film materials is discussed~\cite{Yang_thinfilm}. The possibility of manipulating the transport properties of these materials using a magnetic field offers another interesting avenue to explore, which may be relevant to recent and future experiments~\cite{Ueda_Bfield,tian2015field}.

Second, we will consider the three-dimensional $\beta$-Li$_2$IrO$_3$\cite{takayama2015hyperhoneycomb} and $\gamma$-Li$_2$IrO$_3$\cite{Modic2014ch} compounds, in which the interplay of strong SOC and strong electronic correlations gives rise to highly anisotropic magnetic exchange interactions, including the famous Kitaev interaction. A number of novel ground states arise from a model of these exchange interactions, including a quantum spin liquid phase\cite{Kitaev2003,Mandal2009}, several $q=0$ magnetic orders, multi-q phases and several spiral phases\cite{Lee2014,Kimchi2013,lee2015theory,kimchi2015unified,lee2015two}. We first explain the results of experiments on these systems\cite{takayama2015hyperhoneycomb,Modic2014ch,biffin2014unconventional,biffin2014noncoplanar}. Next, we review the spin liquid phase, the pursuit of which has driven much of the interest in these materials. This phase hosts fascinating properties, including topologically protected surface flat bands and Majorana excitations with a nodal line spectrum\cite{Mandal2009,Lee2014,mandal2014fermions,Modic2014ch,Schaffer2014ts,hermanns2014quantum,mullen2015}. It is also expected to persist to finite temperatures, and a number of possible experiments have been proposed to examine this order\cite{Nasu2014ft,nasu2014vaporization,knolle2014raman,Perreault2015,Smith2015neutron}. After this, we discuss the possible magnetic phases which arise from a classical treatment of two proposed models for the spin exchange interactions in these materials. In particular, we focus on the realization of novel spiral phases that are closely related to those seen in the experiments.

We will close with a discussion of how these systems fit into the greater picture of spin-orbit coupled systems. A number of other important examples that are not discussed in the bulk of this review will be briefly discussed there. In particular, we will mention K-doped Sr$_2$IrO$_4$ and SrIrO$_3$ as possible platforms for superconductivity and a new semi-metal phase. We will also discuss the physics of $\alpha$-RuCl$_3$, which is a prominent example of 4d transition metal oxides with enhanced spin-orbit coupling.

\section{\label{sec:intro} Pyrochlore Iridates}
In this section, we review recent theoretical and experimental progresses
in the study of pyrochlore iridates R$_{2}$Ir$_{2}$O$_{7}$,
where R is a rare-earth element (for example, R=Y, Eu, Sm, Nd, Pr).
Here the Ir$^{4+}$ ions and R are sitting on two inter-penetrating
pyrochlore lattices~\cite{structure_Rao,structure_Maeno}.
These materials enter a magnetic insulator phase at a certain temperature
which becomes lower for larger ionic sizes of the rare-earth element~\cite{MIT}.
In particular, magnetic order is not seen in the stoichiometric Pr$_2$Ir$_2$O$_7$,
which remains metallic down to low temperatures~\cite{PrIridate1,PrIridate2}. Here, the Pr ion has the largest
ionic size among the rare-earth ions in this class of materials.
It has been suggested that different rare-earth ionic sizes affect the effective
band width of J$_{\rm eff}$=1/2 conduction electrons from Ir$^{4+}$ ions
with strong spin-orbit coupling. Then, the varied relative strength of electron
correlation compared to the effective band width provides an
explanation of the insulator-metal crossover~\cite{Pesin}.

Since the energy scales of the spin-orbit coupling, the effective band width, and
the electron correlation are often comparable in these compounds,
various novel correlated topological phases are predicted to occur, such as
a $Z_{2}$ topological insulator~\cite{Pesin,Yang_pyrochlore}, a Weyl semimetal~\cite{Weyl_iridate,weyl,William1,William2}, an axion insulator~\cite{Weyl_iridate,Ara}, and
a topological Mott insulator~\cite{Pesin, TPChoy,Fiete,Imada1,Imada2}, etc.
The bulk phase diagram and the characteristic electronic properties
of the corresponding phases are already reviewed in a recent
pedagogical review article~\cite{William_review}.
In particular, a simple model for the J$_{\rm eff}$=1/2 conduction band with
an on-site Hubbard-$U$ interaction is employed to discuss the emergence of the
quadratic band-touching semi-metal for zero or small $U$, the all-in/all-out (AIAO)
magnetic order in the magnetic insulator phase at large $U$, and
the Weyl semi-metal phase with AIAO at intermediate strength of $U$~\cite{William1,William2,William_review}.
In the current review, we will discuss tremendous progress made in theoretical
and experimental studies of these systems after the initial review article was written.
Here we provide an overview of recent developments, before we discuss
more details later.

First, there exists strong experimental evidence for the AIAO order in the
magnetic insulator phase. The resonant elastic X-ray scattering
experiment on Eu$_2$Ir$_2$O$_7$~\cite{Nakatsuji_xray} and the neutron scattering experiment on Nd$_2$Ir$_2$O$_7$~\cite{Tomiyasu_neutron}
are earlier examples of this. This conclusion is further consolidated by very recent magneto-transport and
magnetization measurements on single crystals of Nd$_2$Ir$_2$O$_7$ in the presence of an external
magnetic field, applied along different crystallographic directions~\cite{Ueda_Bfield,tian2015field}. As discussed later in this
review, the dependence of magneto-transport and magnetization on the direction of
the magnetic field is consistent with the pre-existent AIAO order in the absence of the applied magnetic field.
Moreover, strong evidence for the presence of domain walls between AIAO and AOAI
magnetic domains has been discovered~\cite{Ueda_DW,Ueda_optical,DW_Shen}. A recent inelastic resonant X-ray scattering
experiment on Eu$_2$Ir$_2$O$_7$ reveals excitation spectra that are also consistent with
spin-wave excitations in the AIAO order~\cite{xray2,Eric}.
Secondly, a recent ARPES (angle resolved photoemission) experiment on
a single crystal of Pr$_2$Ir$_2$O$_7$ provides strong evidence for a quadratic
band-touching dispersion in the semi-metallic paramagnetic phase~\cite{QBC_Arpes}.
As to the Weyl semi-metal phase, there has been no direct experimental evidence.
On the other hand, the gapless behavior in the optical conductivity for Rh-doped Nd$_2$Ir$_2$O$_7$~\cite{Ueda_optical}
and the insulator-metal transition observed in Eu$_2$Ir$_2$O$_7$ with hydrostatic pressure~\cite{Tafti}
may offer further opportunities to discover the Weyl fermion phase in these materials.

On the theoretical front, the nature of bulk quantum phase transitions between different
phases and the associated quantum critical behaviors have been intensively studied.
In particular, the roles of the long-range Coulomb interaction in the quadratic band-touching
and Weyl semi-metal phases have been studied~\cite{Moon}. Moreover, the influence of the long-range
Coulomb interaction on the quantum phase transitions and quantum critical points have been investigated~\cite{Savary, Yang_2014}.
We discuss novel non-Fermi liquid and other quantum critical behaviors found in these studies.

Another important development is the understanding of the bulk-boundary correspondence.
The surface state of the Weyl semi-metal phase is characterized by the Fermi arc
that connects two points in the surface Brillouin zone, which are the projections of
a pair of Weyl points in the bulk Brillouin zone~\cite{Weyl_iridate}. A pair of Weyl points are the source/sink
of the Berry flux in momentum space and can be described as
the monopole/anti-monopole of the Berry flux. Any 2D plane in momentum
space, which cuts the line between two bulk Weyl points, would carry a finite Chern number
and acts as a system with integer quantum Hall effect. Hence there should be a gapless
boundary state for such a 2D plane, which corresponds to a gapless point in the surface Brillouin zone.
Connecting all of these gapless points, one obtains the Fermi arc.
When there exist only a pair of Weyl points, one would expect a large anomalous Hall effect
with the Hall conductivity proportional to the distance between two Weyl points~\cite{Ran}.
In the bulk materials, there are other Weyl points related by the cubic crystal
symmetry, and the net anomalous Hall effect is zero~\cite{Ran}.

This suggests that breaking the cubic symmetry of the bulk crystal may be
a profitable avenue to generate a large anomalous Hall effect.
As discussed later in this review, thin films of the pyrochlore iridates can be used to
simulate this physics~\cite{Yang_thinfilm}. Here the surface states of thin films play important roles
in the transport properties of thin films. We discuss possible quantum phase
transitions in the surface states depending on the nature of the boundary.
These ideas can further be extended to physics of magnetic domain walls~\cite{Yamaji,Yamaji2}.
Recently, the fabrication of thin films of Eu$_2$Ir$_2$O$_7$ has been
reported and would be a promising platform to explore possible control of
the surfaces/interfaces/domain-walls in atomic precision.\cite{fujita2015odd}

The last topic we discuss is the magnetic field control of the ground state properties
in pyrochlore iridiates.
In some compounds, the rare-earth ion R (for example, Nd) may carry a magnetic
moment that is much bigger than the Ir moment.
Thus, the application of the external magnetic field can control the R spin configuration,
which, in turn, modulates the Ir magnetic ordering. Since the transport property
of the materials strongly depends on the Ir magnetic configurations,
it is possible to control the transport properties of pyrochlore iridates
via the application of a magnetic field. We discuss this in light of the recent
experiments on single crystals of Nd$_2$Ir$_2$O$_7$~\cite{Ueda_Bfield,tian2015field}.

\subsection{\label{sec:QC} Bulk quantum phase transitions}
\begin{figure*}
\centering
\includegraphics[width=0.8\hsize]{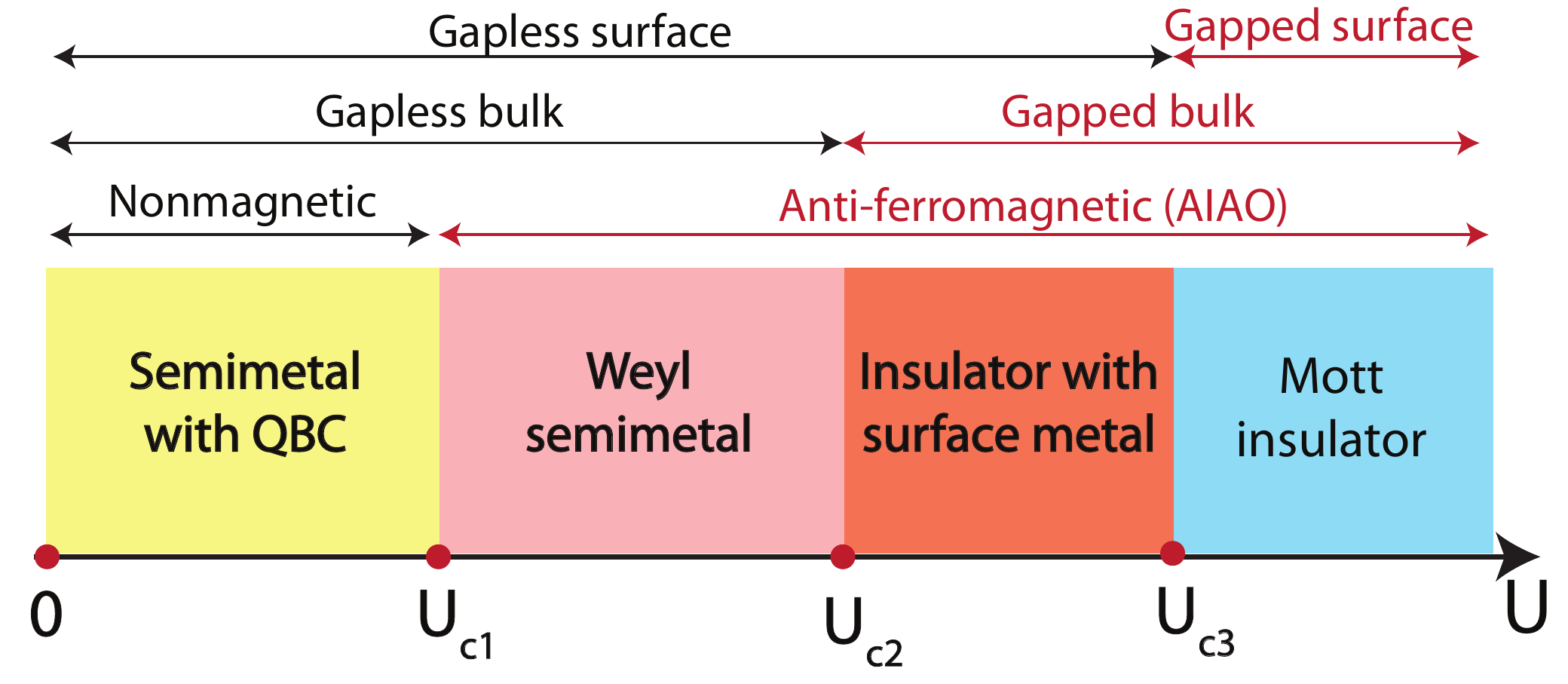}
\caption{
{\bf Generic phase diagram of pyrochlore iridate with antiferromagnetic order, taking into account the property of open surfaces, as a function of
the on-site Coulomb interaction $U$.}
$U_{c1}$ indicates the onset of the all-in all-out (AIAO) antiferromagnetic ordering.
$U_{c2}$ denotes the critical point where the bulk-gap opens via the pair-annihilation of Weyl points.
Depending on surface termination, a bulk insulator with metallic surfaces can arise when $U_{c2}<U<U_{c3}$.
For example, when the surface is perpendicular to the [111] direction and
given by the Kagome layer, the surface metallic states appear even though the bulk gap is finite when $U_{c2} < U < U_{c3}$.
On the other hand, in the case of triangular lattice termination, the origin of the surface states is distinct
so that the surface states are pushed away from the Fermi level.
$U_{c3}$ describes the critical point where surface metallic states completely disappear.
}
\label{fig:phasediagrams}
\end{figure*}

A generic phase diagram of pyrochlore iridates as a function of the on-site Coulomb interaction $U$ is shown in Fig.~\ref{fig:phasediagrams}.
There are four possible phases separated by three quantum critical points $U_{c1}$, $U_{c2}$, and $U_{c3}$.
At first, when $0<U<U_{c1}$,
the ground state is a non-magnetic semi-metal with a quadratic band crossing (QBC)
at the Brillouin zone center.
Due to the peculiar energy dispersion near the Fermi point in this semimetal phase,
the long-range Coulomb interaction is a relevant perturbation,
which drives the system to a non-Fermi liquid phase dubbed the
Luttinger-Abrikosov-Beneslavskii (LAB) phase~\cite{Moon}.

The first critical point $U_{c1}$ indicates
the onset of magnetic ordering.
The anti-ferromagnetic ground state possesses the so-called
all-in/all-out (AIAO) spin ordering in which all four spins at the vertices of
a tetrahedron point into or out of its center~\cite{Weyl_iridate,William1,Arima}.
Naively, since the AIAO spin ordering can be described by an Ising transition in (3+1) dimensions,
the nature of the quantum critical point at $U_{c1}$ seems to be trivial.
However, once one takes into account the long-range Coulomb interaction,
the non-Fermi liquid physics associated with the LAB phase sets in,
which, in turn, makes the magnetic transition unconventional~\cite{Savary}.

Once the AIAO spin ordering develops for $U_{c1}<U<U_{c2}$,
the semimetal with QBC turns into a Weyl semimetal
having eight Weyl points along the [111] and symmetry-related directions.
The Weyl points only come in pairs, each carrying opposite quantized monopole
charge in momentum space. Hence a pair of Weyl points act as a source and
a sink of Berry gauge flux, respectively. This is the fundamental origin of various
topological phenomena such as anomalous Hall effect, negative magneto-resistance associated
with the chiral anomaly, etc.
Since the monopole charges guarantee the stability of a pair of Weyl points, as long as
the interaction strength is not so strong, a pair of Weyl points simply move in momentum
space along the [111] or its symmetry-related directions as $U$ increases.
When a pair of Weyl points with opposite monopole charges merge at the same momentum,
which happens at the critical point $U_{c2}$, two Weyl points can be pair-annihilated, and
the bulk energy gap is opened for $U>U_{c2}$.

At the critical point $U_{c2}$, the long-range Coulomb interaction brings about
intriguing quantum critical phenomena again.
Contrary to the critical point $U_{c1}$ where the time-reversal symmetry
of the system is broken due to the magnetic ordering,
the critical point $U_{c2}$ indicates a Lifshitz-type topological phase transition,
and the symmetry of the system is preserved.
When two Weyl points having opposite monopole charges merge at the same momentum,
the energy dispersion near the gap-closing point becomes
highly anisotropic. Namely,
the dispersion becomes quadratic along one direction
and linear along the other two orthogonal directions.
The quantum fluctuation of electrons with anisotropic dispersion
screens the long-range Coulomb interaction, leading to an anisotropic, but
still long-ranged, interaction.
Such an anisotropic screening makes the Coulomb interaction
irrelevant in the low energy limit. As a result, the electrons
remain as well-defined quasiparticles~\cite{Yang_2014}.

Depending on the nature of the surface layer, gapless Fermi surface states can arise
at the boundary to the vacuum even after the bulk gap opens at $U_{c2}$.
For example, when the boundary layer in the [111] direction is given by the Kagome lattice,
it can be shown that
Fermi surfaces arise at the boundary for a finite window of $U_{c2} < U < U_{c3}$.
In the context of surface states, the critical point $U_{c2}$
marks the Lifshitz transition at which the Fermi surface topology
changes from an open arc to a closed loop.
Such surface states with a closed Fermi loop topology generate
an interesting phase with metallic surfaces and
gapped bulk states. Such a surface metallic phase
can induce nontrivial interfacial conductance, and
may be used to engineer new types of topological phases in thin films.

Finally, when $U>U_{c3}$, both bulk and surface states are fully gapped, and
the system turns into a conventional Mott insulator.
In the following, we review the intriguing properties of the system in each region of the phase
diagram and the associated quantum critical points.

\subsubsection{When $0<U<U_{c1}$: quantum critical Luttinger-Abrikosov-Beneslavskii (LAB) phase}

When the electron-electron interaction is weak, the non-magnetic ground state of pyrochlore iridates can be described by
a semimetal having a quadratic band crossing (QBC) point at the zone center.
Although the presence of the point-like Fermi surface is not generically guaranteed and small electron/hole pockets
can be developed, recent first-principles calculations~\cite{QBC_DFT} and ARPES measurement~\cite{QBC_Arpes} consistently show that
there are several pyrochlore iridate compounds having
the quadratic band crossing point (QBC) at the Fermi energy.
Moreover, the aperiodic quantum oscillation signals observed in the paramagnetic
state of Pr$_2$Ir$_2$O$_7$, both in experiment~\cite{PrIridate1,PrIridate2,SdH_exp} and in theory~\cite{SdH_theory},
provide additional support for the presence of the QBC.

The quadratic dispersion near the zone center can be understood
by considering the cubic symmetry of the system.
The relevant effective $k\cdot p$ Hamiltonian,
which is generally called the Luttinger Hamiltonian,
has the following structure with three mass parameters $m_{0}$, $m$, $M_{c}$~\cite{Luttinger1,Luttinger2},
\begin{eqnarray}\label{eqn:luttinger}
H(\textbf{k})&=&\frac{k^{2}}{2m_{0}}+\frac{d_{a}(\textbf{k})\Gamma_{a}}{2m}+\frac{1}{2M_{c}}\{d_{4}(\textbf{k})\Gamma_{4}+d_{5}(\textbf{k})\Gamma_{5}\},
\end{eqnarray}
where the summation over repeated indices is implied.
In the Hamiltonian, $\Gamma_{1,2,3,4,5}$ are $4\times 4$ Gamma matrices satisfying
$\{\Gamma_{a},\Gamma_{b}\}=2\delta_{ab}$ and
\begin{align}
d_{1}(\bm{k})&=\sqrt{3}k_{y}k_{z},~d_{2}(\bm{k})=\sqrt{3}k_{z}k_{x},~d_{3}(\bm{k})=\sqrt{3}k_{x}k_{y},
\nonumber\\
d_{4}(\bm{k})&=\frac{\sqrt{3}}{2}(k_{x}^{2}-k_{y}^{2}),~d_{5}(\bm{k})=\frac{1}{2}(2k_{z}^{2}-k_{x}^{2}-k_{y}^{2}).
\end{align}
From the Hamiltonian in equation (\ref{eqn:luttinger}), we obtain the energy eigenvalues
\begin{align}
E_{\pm}(\bm{k})=\frac{k^{2}}{2m_{0}}\pm\sqrt{c_{a}d_{a}(\bm{k})d_{a}(\bm{k})}
\end{align}
where $c_{a=1,2,3}=1/(2m)$ and $c_{a=4,5}=1/(2m)+1/(2M_{c})$.
The quadratic dispersion near the zone center with $\bm{k}=0$ is evident.

Due to the vanishing density of states at the Fermi level,
the screening of the long-range Coulomb interaction is ineffective in semimetals.
As a result, the Coulomb interaction retains its long-range nature.
Notice that the quadratic energy dispersion supports the low-energy density
of states $D(E)\propto \sqrt{E}$, which is enhanced compared to
the case of Dirac semimetals which have a linear energy dispersion where $D(E)\propto E^{2}$.
It is well known that the long-range Coulomb interaction is marginal in Dirac semimetals,
hence it becomes a strongly relevant perturbation in the semimetal with QBC.
The renormalization group analysis shows that there exists a strong coupling fixed point,
where the long-range Coulomb interaction becomes a shorter, but still long-range, interaction while
the low energy electrons are characterized by non-Fermi liquid behavior. The non-Fermi liquid phase
at the interacting fixed point is called the Luttinger-Abrikosov-Beneslavskii (LAB) phase~\cite{Moon,Abrikosov1,Abrikosov2}.
Another interesting point is that the cubic anisotropy becomes
irrelevant at the interacting fixed point and hence the LAB phase is an isotropic phase.
As a quantum critical non-Fermi liquid phase,
the LAB phase exhibits universal power-law exponents in various physical observables
such as conductivity, specific heat, nonlinear susceptibility, and the magnetic Gruneisen parameter.
In particular, it is shown that the anomalous Hall conductivity scales as
$\sigma_{xy} \sim {e^2 \over h} \sqrt{M}$ where $M$ is a magnetization or an effective magnetic field.
This means a very small magnetization is enough to generate a relatively large anomalous Hall signal.
This may be consistent with the experimental results on Pr$_2$Ir$_2$O$_7$,
where a large anomalous Hall effect with vanishingly small magnetization has
been observed~\cite{PrIridate1,PrIridate2,2015_Nakatsuji,2013Sungbin}.

\subsubsection{At $U=U_{c1}$: Magnetic quantum critical point associated with all-in/all-out (AIAO) spin ordering}

$U=U_{c1}$ marks the quantum critical point describing
the onset of the AIAO antiferromagnetic ordering.
If we neglect the long-range Coulomb interaction,
the magnetic transition can be described by
the following action
\begin{align}
S=&\int d^{3}xd\tau\psi^{\dag}\left(\partial_{\tau}+H_{L}(-i\nabla)+gM\phi\right)\psi
\nonumber\\
&+\int d^{3}xd\tau\frac{1}{2}\left[(\nabla\phi)^{2}+(\partial_{\tau}\phi)^{2}+r\phi^{2} \right],
\end{align}
where $\phi$ indicates the order parameter field describing AIAO ordering,
$g$ is the coupling of electrons to the order parameter, and $M=i\Gamma_{4}\Gamma_{5}$ is
the $4\times 4$ matrix relevant to the AIAO order.
Since the quadratic dispersion of electrons implies the dynamical exponent $z=2$,
the inequality $d+z>4$ may seem to hold in this case.
Then according to the standard Hertz-Millis theory,
the critical theory is above the upper critical dimension,
thus the magnetic transition would be mean-field-like.

However, one can easily see that this naive expectation is invalid even in the mean-field treatment.
For instance, if the fermions are integrated out, one obtains a non-analytic term $|\phi|^{5/2}$,
which overwhelms the conventional $\phi^{4}$ term.
In fact, simple power counting shows that the fermion-order parameter coupling $g$
is marginal in the RG sense, thus the theory should be more carefully treated.

To understand the nature of magnetic fluctuations,
one can perform a large-$N$ expansion by introducing $N$ copies
of fermions~\cite{Savary}. The nontrivial nature of magnetic fluctuations
can be seen by computing the boson (order parameter) self-energy in the $N=\infty$ limit.
\begin{align}
\Sigma_{\phi}(\omega_{n},\bm{k})=r_{c}+g^{2}\left( |\ln\frac{c_{1}}{c_{2}}||\bm{k}|f({\bm{\hat{k}}}) +C_{\phi}\sqrt{|\omega_{n}|} \right)
\end{align}
where $r_{c}\sim g^{2}\Lambda$ with the momentum cutoff $\Lambda$, and the constant $C_{\phi}\approx 1.33$.
One can see that the self-energy is much larger than the bare terms $\bm{k}^{2}$ and $\omega_{n}^{2}$
in the small momentum and frequency limits, and thus dominates the boson propagator.
Computing the electron self-energy and vertex function by taking this dressed boson propagator,
the RG flow of $c_{1}/c_{2}$ and $c_{0}/c_{1}$ can be obtained. Here the bare value of $c_0$
is given by $1/m_0$ or the coefficient of the isotropic part of the kinetic Hamiltonian.
It turns out that $c_{1}/c_{2}$ is an irrelevant parameter indicating the extreme cubic anisotropy
of the interacting system.
On the other hand, $c_{0}/c_{1}$ is a relevant parameter, thus $c_{0}$ dominates
over the $c_{1}$ in the low energy limit.
It can be shown that the quadratic band crossing can be exactly at the Fermi level
only if $c_{0}<c_{1}/\sqrt{6}$.
Once $c_{0}$ becomes bigger than $c_{1}/\sqrt{6}$,
Fermi surfaces start to develop, rendering the critical theory invalid.
Namely, the transition from the semimetal with QBC to a Weyl semimetal is interrupted
by a Lifshitz transition developing Fermi surfaces.

Interestingly, the long-range Coulomb interaction plays a nontrivial
role in stabilizing the quantum critical point.
The full theory, including the long-range Coulomb interaction and
order parameter fluctuations, is given by
\begin{align}
S=&\int d^{3}xd\tau\psi^{\dag}\left(\partial_{\tau}+H_{L}(-i\nabla)+ie\varphi+gM\phi\right)\psi
\nonumber\\
&+\int d^{3}xd\tau\frac{1}{2}\left[(\nabla\varphi)^{2}+(\nabla\phi)^{2}+(\partial_{\tau}\phi)^{2}+r\phi^{2} \right],
\end{align}
where $\varphi$ describes the electrostatic field mediating the Coulomb interaction.
To determine the RG flow of $c_{1}/c_{2}$ and $c_{0}/c_{1}$,
the coupling of electrons to the Coulomb potential $\varphi$ and order parameter
fluctuations $\phi$ should be considered simultaneously.
The resulting RG equations show that both $c_{1}/c_{2}$ and $c_{0}/c_{1}$
are irrelevant, and
there is a hierarchy of $c_{0}\ll c_{1}\ll c_{2}$.
Therefore, the tendency of the Coulomb interaction to suppress the particle-hole
asymmetry makes the critical theory stable.

The resulting quantum critical point displays
various novel features including highly unconventional critical exponents
and emergent spatial anisotropy.
In particular, unlike most classical and quantum critical points,
spatial rotational symmetry is strongly broken, which may lead to
spiky Fermi surfaces if the system is close to the undoped quantum
critical point and/or the doping is sufficiently small.

\subsubsection{At $U=U_{c2}$: quantum critical point of Weyl semimetal-insulator transitions}
When $U_{c1}<U<U_{c2}$, the Weyl points which are initially created
at the zone center move along the [111] or its symmetry-related directions
as $U$ increases. At $U_{c2}$, all Weyl points hit the BZ boundary
and pairs of Weyl points having opposite monopole charges merge at the same momentum.
The merging of two Weyl points can be described by the following Hamiltonian,
\begin{align}
H=vk_{x}\tau_{x}+vk_{y}\tau_{y}+(Ak_{z}^{2}+m)\tau_{z},
\end{align}
where $v$ is the velocity in the $k_{x}$ and $k_{y}$ direction,
and $1/(2A)$ is the effective mass along the $k_{z}$ direction,
which we assume to be positive.
Then depending on the sign of the parameter $m$,
we have either a Weyl semimetal ($m<0$) with two Weyl points on the $k_{z}$ axis,
or a gapped insulator ($m>0$).
At the critical point with $m=0$, corresponding to $U_{c2}$ in pyrochlore iridates,
the two Weyl points with the opposite monopole charges merge at the same momentum $\bm{k}=0$.
The effective Hamiltonian at the critical point is
\begin{align}
H_{c2}=vk_{x}\tau_{x}+vk_{y}\tau_{y}+Ak_{z}^{2}\tau_{z},
\end{align}
where the electron dispersion is linear in two directions and is quadratic
in the third direction, representing an anisotropic Weyl fermion (AWF).
The appearance of a gapless point with an anisotropic dispersion
is a generic property of the quantum critical point for semimetal-insulator transitions~\cite{Murakami_2007,Murakami_2008,Yang_2013}.

Quantum fluctuations of AWF screen the Coulomb potential anisotropically,
which is reflected in the polarization function
\begin{align}
\Pi(\bm{q})\sim b_{\perp}(q_{1}^{2}+q_{2}^{2})^{3/4}+ b_{3}q_{3}^{2},
\end{align}
where $b_{\perp}$ and $b_{3}$ are constants.
One can clearly see that the polarization function has unusual momentum dependence
along the linear-dispersion direction whereas the momentum dependence of $\Pi(\bm{q})$
has the conventional quadratic form along the quadratic-dispersion direction.
In the $q_{1}$ and $q_{2}$ direction, the boson self-energy represented by
the polarization function dominates over the bare Coulomb potential in the small momentum region,
which fundamentally changes the properties of the Coulomb interaction~\cite{Yang_2014,Abrikosov3}.
For instance, the screened Coulomb interaction has
the following coordinate dependence in real space,
\begin{align}
V(z=0)\sim \frac{1}{(x^2+y^2)^{5/8}},~~V(x=y=0)\sim \frac{1}{|z|^{5/3}}.
\end{align}
Notice that the Coulomb interaction develops spatial anisotropy while maintaining its long-ranged form,
which is clearly distinct from the conventional Thomas-Fermi screening and
the logarithmic screening in Dirac semimetals with linear dispersion.
Such a screened Coulomb interaction, however, can be shown to be an irrelevant perturbation
at the interacting fixed point and the AWF remain well-defined quasiparticles.

\subsection{\label{sec:thinfilm} Surface quantum phase transitions}

\begin{figure*}
\centering
\includegraphics[width=0.8\hsize]{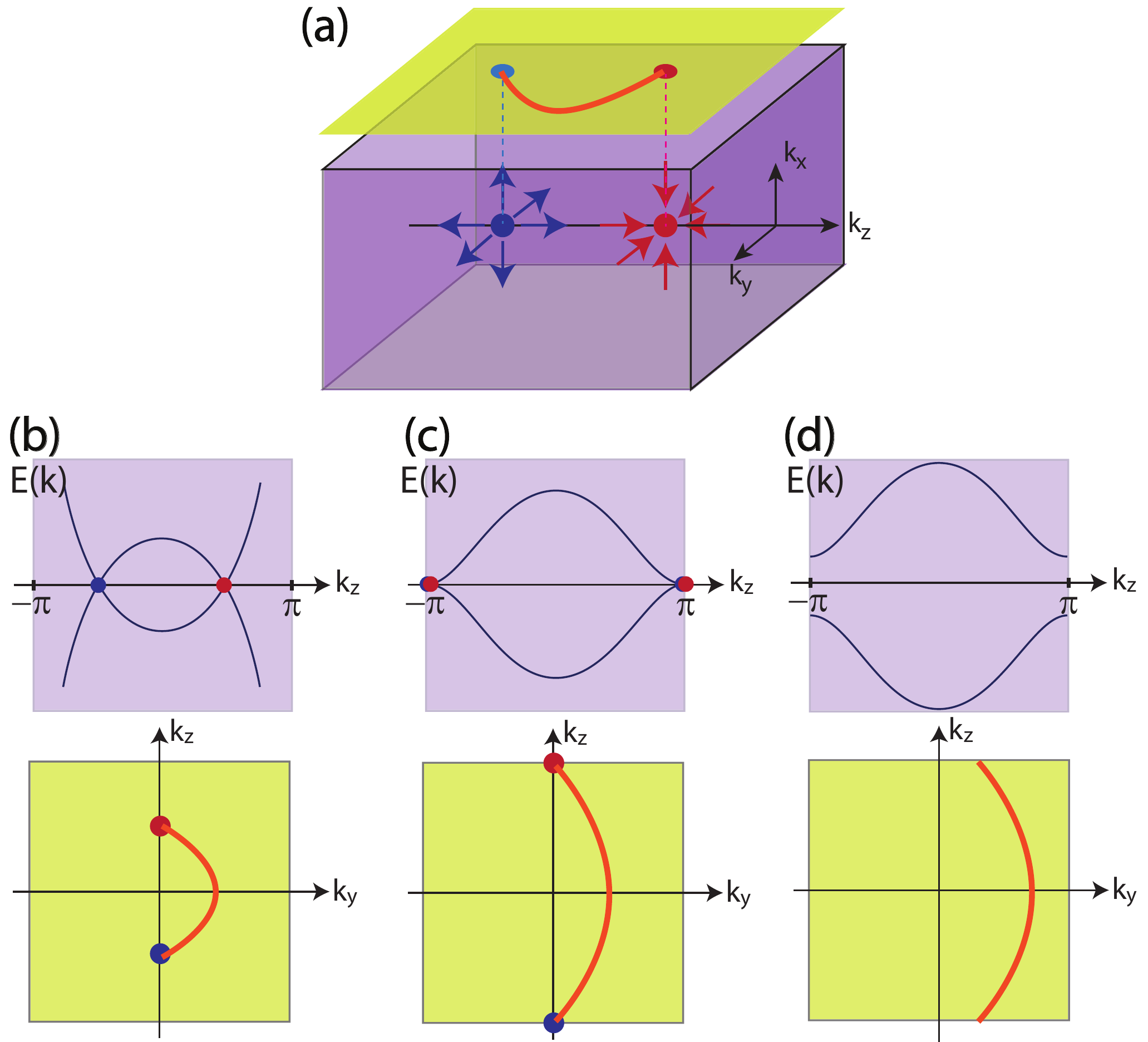}
\caption{
{\bf Evolution of the bulk and surface spectra of a simple toy model ferromagnet with two bulk Weyl points.}
(a) Location of the two bulk Weyl points with opposite monopole charges and the shape of the corresponding
surface Fermi arc.
(b) The bulk band structure and the associated Fermi arc on the surface BZ
when the system is in the Weyl semimetal phase.
There exists a Fermi arc on each surface of the system as long as two bulk Weyl points are not projected
on top of each other in the surface BZ.
(c) The bulk and surface electronic structures at the quantum critical point, where the pair-annihilation of
two Weyl points happens at the BZ boundary.
There is a Lifshitz transition in the surface spectrum.
(d) The case when the bulk gap is fully opened. The zero energy contour of the Fermi surface in the surface BZ
forms a closed loop since the system is periodic in the $k_{z}$ direction.
Notice that, in this model with ferromagnetism, the system has a nonzero Chern vector, which guarantees the presence of surface states.
In particular, when the bulk gap is fully opened, the system becomes a 3D quantum Hall insulator, which should carry a closed Fermi loop
on the surface as long as the Chern vector is not parallel to the surface normal direction.
}
\label{fig:Weyl_FM}
\end{figure*}

Up to now, we have reviewed the properties of bulk states.
Now let us turn to the physics associated with the surface or the interface of the system.
One intriguing property of Weyl semimetals is the presence of surface states, dubbed the Fermi arc.
The Fermi arc originates from the fact that each Weyl point behaves like
a fictitious magnetic monopole producing Berry curvature in the momentum space~\cite{Weyl_iridate}.

Before we describe the relation between bulk Weyl points and
the corresponding surface spectrum in pyrochlore iridates,
let us consider a simple toy model
composed of two Weyl points at the momenta $\bm{k}=(0,0,\pm k_{W})$ with the monopole charge $\pm 1$, respectively,
as shown in Fig.~\ref{fig:Weyl_FM}.
Such a band structure can occur in ferromagnets when an accidental band crossing happens
between the valence and conduction bands~\cite{Ran,HgCrSe,Burkov}.
It can be shown that, for a given $k_{z}$, a two-dimensional (2D) plane perpendicular to the $k_{z}$ axis
can be considered as a 2D quantum Hall insulator if $k_{z}\in(-k_{W},k_{W})$.
Otherwise, it should be a 2D trivial insulator.
This suggests that the integral of the Berry curvature produced by the Weyl points
leads to a finite anomalous Hall conductivity of $\sigma_{xy}=\frac{e^{2}}{ha_{z}}\frac{2 k_{W}}{2\pi}$,
where $a_{z}$ indicates the lattice constant along the $z$ direction~\cite{Ran}.
Moreover, the presence of quantum Hall layers naturally leads to chiral surface states
on the boundary. Connecting the Fermi point of each chiral edge state in a layer with the given $k_{z}\in(-k_{W},k_{W})$,
we obtain an open Fermi surface, i.e. a Fermi arc. (See Fig.~\ref{fig:Weyl_FM} (a,b).)
Since a Fermi arc exists between two momenta
representing the projection of bulk Weyl points to the surface BZ,
the size of the Fermi arc gets larger as the bulk Weyl points approach
the BZ boundary.
Finally, when the two bulk Weyl points merge at the BZ boundary, the bulk gap is fully opened
and the relevant surface state undergoes a Lifshitz transition,
leading to a closed surface state. (See Fig.~\ref{fig:Weyl_FM} (c,d).)
In this case, the bulk quantum critical point
where Weyl points pair-annihilate corresponds to
a Lifshitz transition point of surface states where an open Fermi surface becomes
a closed Fermi surface.

\begin{figure*}
\centering
\includegraphics[width=0.8\hsize]{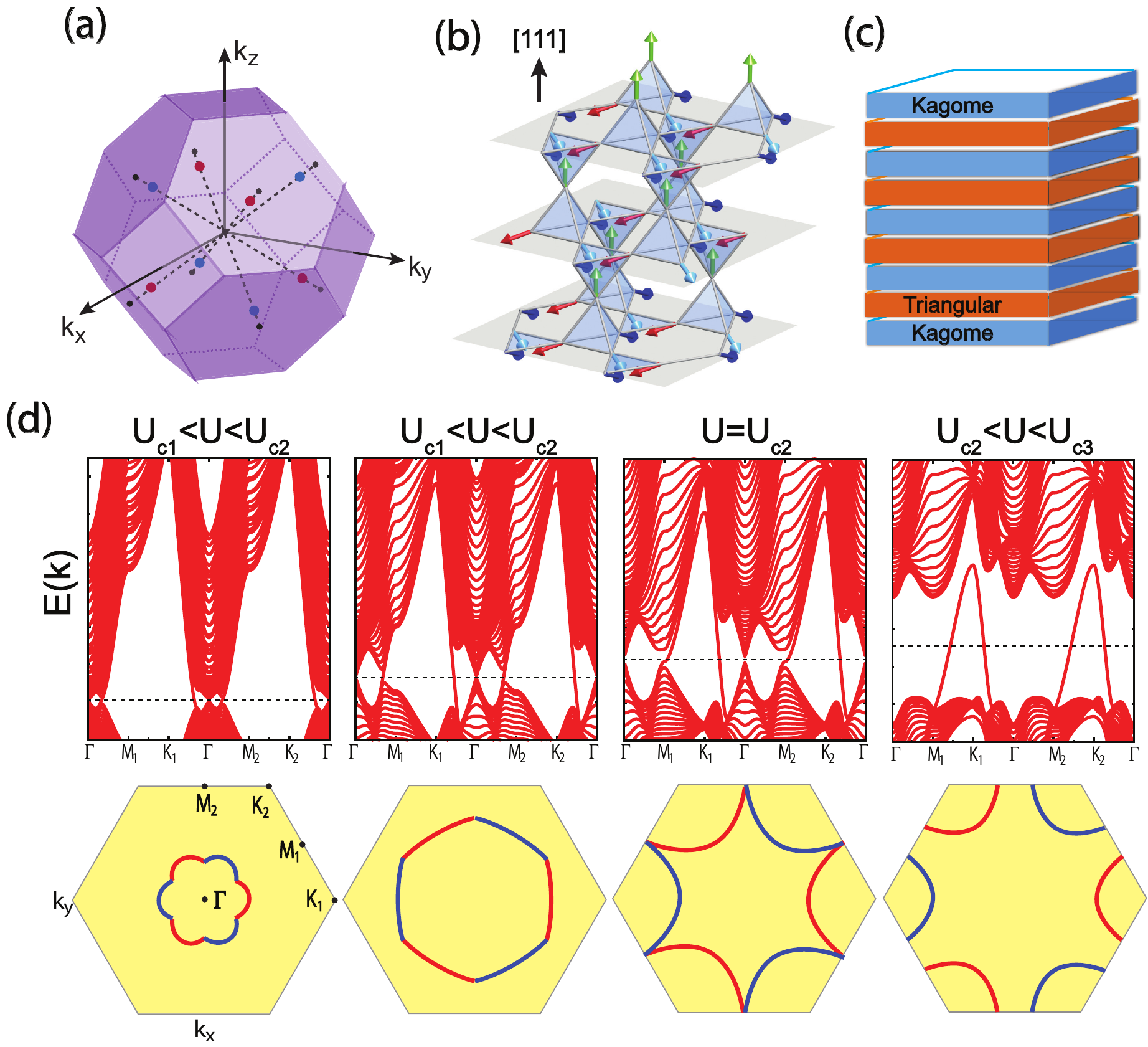}
\caption{
{\bf Evolution of the bulk and surface spectra
of the pyrochlore iridate antiferromagnet with the kagome layer on both the top and the bottom surfaces.}
(a) Locations of 8 Weyl points in the bulk Brilluin zone.
(b, c) Structure of the [111] thin film having the kagome lattice on both the top and bottom layers.
Here we are considering the film composed of $N_{b}$=20 bilayers (triangular and Kagome layers form a bi-layer unit),
plus an additional Kagome layer on the top.
(d) The bulk band structure (upper panels) and the Fermi surface shape of surface states (lower panels).
The dotted line in the top panels indicates the Fermi energy.
In lower panels, the red (blue) lines indicate the surface states at $E_{F}$
localized on the top (bottom) surfaces and there is no surface state in the yellow region.
When $U_{c1}<U<U_{c2}$, the size of the Fermi surface increases as $U$ grows.
At $U=U_{c2}$, the Fermi surface shows Lifshitz transition at which
an open Fermi surface transforms to a closed Fermi surface.
Finally, when $U>U_{c2}$, the Fermi surface is composed
of two isolated Fermi loops encircling two corners of the surface
Brillouin zone. As $U$ increases further,
the size of the Fermi circle reduces, and the surface state eventually disappears at $U=U_{c3}$.
Thus the system becomes a trivial gapped insulator when $U>U_{c3}$.
}
\label{fig:thinfilm}
\end{figure*}

Now let us consider the surface states of the Weyl semimetal in pyrochlore iridate antiferromagnets.
In the simplest model of pyrochlore iridates, due to the cubic symmetry, there are four pairs of Weyl points
in the system, which are aligned along the [111] and symmetry-related directions~\cite{William1}
(In the first-principle numerical computations,
the positions of the Weyl points are off the high symmetry axis due to the lower symmetry, and there exist
three times more Weyl points).
Because of this, the surface
of the Weyl semimetal in pyrochlore iridate antiferromagnets
supports multiple Fermi arcs.
For instance, for a finite slab (or a thin film)
with the surface normal vector along the $[111]$ direction,
the relevant surface spectrum is shown in Fig.~\ref{fig:thinfilm}.
Note that the pyrochlore lattice can be regarded as alternating layers of
Kagome and triangular lattices along the $[111]$ direction.
Here we consider the case when the boundary layers (both at the top and bottom)
to the vacuum is given by the Kagome lattice.
Once a [111] surface is introduced, two Weyl points
in the [111] direction are projected onto the same momentum, i.e., the center of
the surface Brillouin zone whereas
the other six Weyl points are distributed symmetrically around
the BZ center~\cite{Yang_thinfilm}. Therefore only three Fermi arcs, one per pair of Weyl points, can be found
on the surface BZ as shown in Fig.~\ref{fig:thinfilm} (d).
Since the separation between Weyl points increases as $U$ gets bigger,
the size of the Fermi arcs also grow.
At $U=U_{c2}$ where all Weyl points hit the bulk BZ boundary and pair-annihilate,
three Fermi arcs touch at the surface BZ boundary, signalling a Lifshitz transition
from open Fermi surfaces to closed Fermi loops.
As $U$ increases further, the size of the closed Fermi loop
at each corner of the surface BZ reduces, and eventually
the surface states disappear when $U>U_{c3}$.

It is worth noting that the surface states do not disappear immediately
after the opening of the bulk gap at $U=U_{c2}$.
Until the surface state disappears through another Lifshitz transition at $U=U_{c3}$,
an interesting intermediate state, in which bulk states are gapped while the surface states are gapless,
can be realized in the interval of $U_{c2}<U<U_{c3}$.
Though the closed Fermi surface does not have topological stability,
in the sense that it can disappear when $U$ is increased continuously,
it still preserves the chiral nature of the Fermi arcs~\cite{Yang_thinfilm}.
Namely, if there is a surface state at the momentum $\bm{k}$ on the top surface,
the other surface state carrying the momentum $-\bm{k}$ always exists on the bottom surface.
Such a spatial separation of the states at the momentum $\bm{k}$ and $-\bm{k}$
is a hallmark of chiral states; thus, the closed Fermi surface on the surface
has extra stability against back-scattering.
Here we note that the presence of the intermediate phase in $U_{c2}<U<U_{c3}$ depends on the nature
of the boundary to the vacuum. For example, this regime does not exist when the boundary to the vacuum
is given by the triangular lattice both at the top and bottom surfaces.
Hence it may be possible to design interesting thin film or domain wall structures using appropriate
boundaries in the regime $U_{c2}<U<U_{c3}$. This can bring about various novel topological phenomena
that are not possible in bulk materials, as described in the following.

\begin{figure*}[t]
\centering
\includegraphics[width=0.8\hsize]{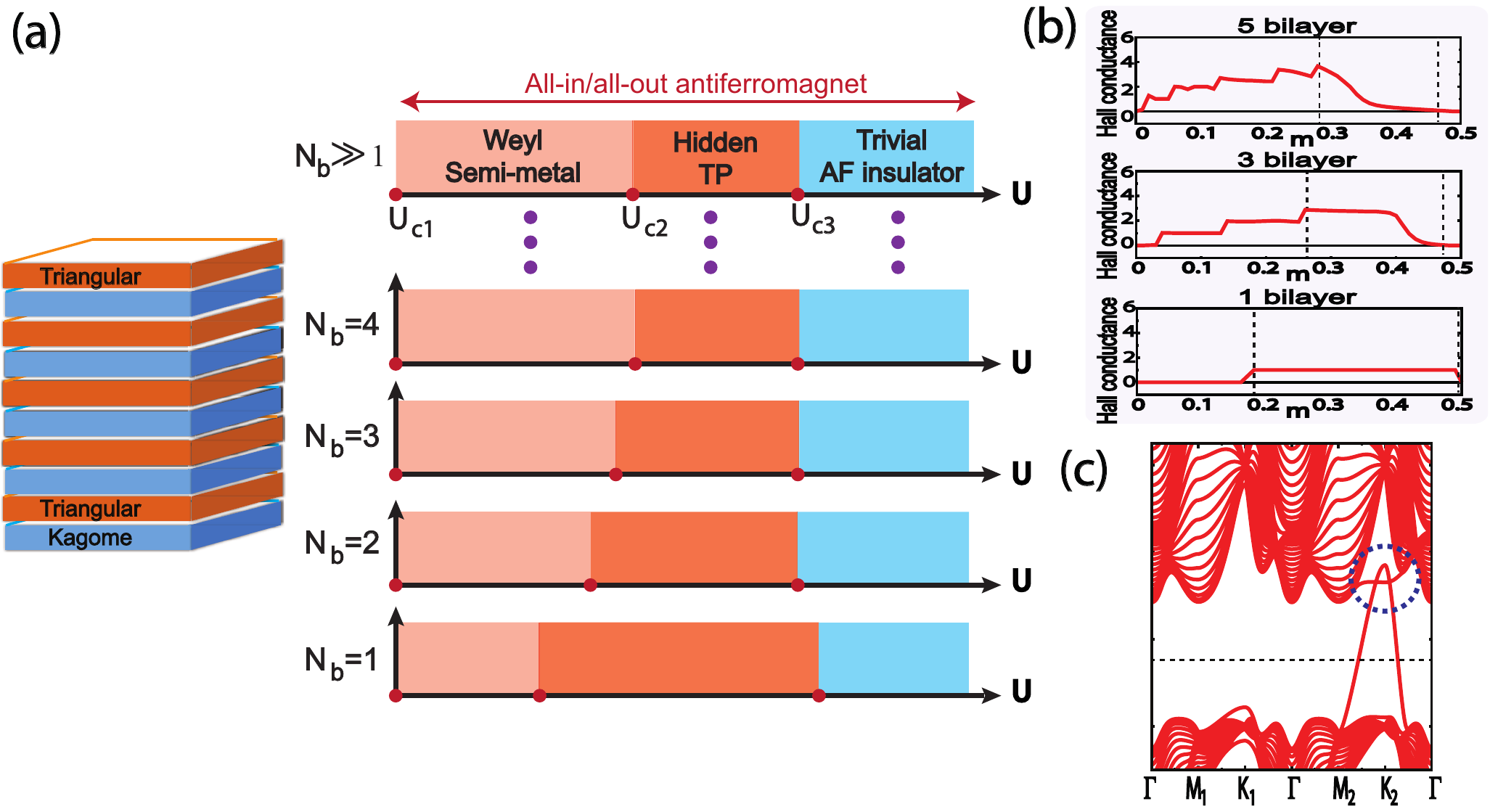}
\caption{
{\bf Dimensional crossover for [111] films of pyrochlore iridate antiferromagnets.}
({\bf a}) Evolution of the phase diagram of [111] thin films, terminated by a triangular layer on one surface
and a Kagome layer on the other surface, as the thickness of films increases.
Here $N_{b}$ denotes the number
of bilayers and neighboring Kagome and triangular layers constitute a unit bilayer.
In the bulk limit of $N_{b}\gg 1$, the bulk gap opens at $U=U_{c2}$.
When $U_{c2}<U<U_{c3}$, the conduction and valence bands are coupled due to the crossing of two surface states as shown in ({\bf c}).
The surface states completely disappear at $U=U_{c3}$.
In the pink region of the phase diagram, the Hall conductance $G_{xy}$ is maximum at the left phase boundary
whereas $G_{xy}$ vanishes at the right phase boundary.
({\bf b}) Behavior of $G_{xy}$ in different thin film configurations. Two dotted vertical lines in each film case
indicate the corresponding $U_{c2}$ and $U_{c3}$, respectively.
({\bf c}) The band structure of a thick film with $N_{b}=20$ when $U_{c2}<U<U_{c3}$.
The dotted circle marks the crossing of the two surface states localized on the top and bottom surfaces, respectively.
} \label{fig:hallconductance}
\end{figure*}

\subsubsection{Emergent topological properties of thin films}

Manipulation of thin films is a promising way to unveil the intriguing topological properties
of the bulk Weyl semimetal proposed in pyrochlore iridate antiferromagnets~\cite{Yang_thinfilm,Fiete_film,Bergholtz_film,Fiete_film2,Fiete_film3}.
Though a bulk Weyl point
is a potential source of an anomalous Hall effect,
the anomalous Hall currents generated by different Weyl points
are canceled in the bulk system due to the cubic symmetry of the lattice~\cite{Ran}.
However, once
the system is prepared in a thin film form, which intrinsically breaks the cubic symmetry,
the incomplete cancellation of
the Berry curvature can give rise to a finite anomalous Hall currents~\cite{Yang_thinfilm}.
Moreover, since different choices of the boundary/surface would strongly influence the
nature of the electronic states, the interplay between different surface states can trigger
unique topological properties in thin films, which cannot be expected from a simple extrapolation
of the phase diagram of bulk systems with translational invariance.

Viewed along the [111] direction, a pyrochlore lattice is composed
of two-dimensional (2D) kagome and triangular lattice layers stacked alternatively.
Due to this peculiar lattice structure, the [111] direction is a natural direction
to grow thin films.
To reveal the topological properties of [111] thin films,
let us consider the Hall conductance $G_{xy}$ as a function
of the thickness of the film.
Here the thickness of a thin film is controlled by changing
the number of bilayers ($N_{b}$) where
a bilayer is composed of a kagome layer and its neighboring triangular layer.
Thus, in each film, one surface is terminated by a triangular layer whereas the other surface is terminated
by a Kagome layer.

As shown in Fig.~\ref{fig:hallconductance},
the phase diagram of such thin films has an unexpectedly rich structure.
At first, when $U_{c1}<U<U_{c2}$, $G_{xy}$
increases monotonously
as $U$ increases, and eventually reaches the maximum value
at $U=U_{c2}$.
Remarkably, the finite AHE can be observed even beyond the bulk quantum
critical point ($U_{c2}$).
Since bulk states are fully gapped for $U>U_{c2}$,
the variation of $G_{xy}$ is possible only if the film supports intrinsic surface states
connecting the bulk valence and conduction bands.
Here the surface states existing for $U_{c2}<U<U_{c3}$ play the crucial role.
The band structure of a film with $N_{b}=20$ when $U_{c2}<U<U_{c3}$ is shown in Fig.~\ref{fig:hallconductance} (c).
One can clearly notice the presence of the bulk gap in most parts of the BZ while
the conduction and valence bands are still connected due to the crossing of two surface states.
Among these two states touching each other, one state connected to the bulk valence band comes from
the surface terminated by a Kagome layer whereas the other state connected to the conduction band
is localized on the other surface terminated by a triangular layer.
The crossing of two surfaces states localized on the top and bottom layers of the film
generates large anomalous Hall currents, which is the origin of the finite Hall conductance
when $U_{c2}<U<U_{c3}$.
When $U>U_{c3}$, the two surface states decouple, and the film becomes topologically trivial
with vanishing $G_{xy}$.

As described above, thin films of pyrochlore iridates
offer unique opportunities in various aspects.
(i) One can show that the AHE ($G_{xy}$) is quantized when $N_{b}=1$ or $2$.
This would be the first realistic antiferromagnet showing
the quantized AHE due to the scalar spin chirality
{\it without} the uniform magnetization.
When $N_b > 2$, the AHE is no longer quantized due to
complex structure of the electronic states, but becomes large and finite.
(ii) The antiferromagnetic phase has two degenerate ground states, i.e., the AIAO state and its time-reversed partner.
Since the Hall conductances of these two phases have
the opposite sign, metallic conducting channels can appear at domain walls.
(iii) The bulk-surface correspondence of the hidden topological phase in $U_{c2} < U < U_{c3}$
belongs to a new class, i.e.,
the surface states carry a large net Berry flux (or a large Chern number), which in principle can be proportional to
the thickness $L_{z}=N_{b}a_{z}$ of the film, while the 3D bulk system ($L_z \to \infty$)
is not topological and has Chern number zero.
This apparently contradicting behavior is explained by the fact that
exponentially small overlap, i.e., $\sim \exp(-L_z/\xi)$
($\xi \sim ta/E_G$ is the correlation length determined by
the transfer integral $t$, the band gap $E_G$ and the lattice constant $a$),
between the surface states on the top and bottom produces the finite Chern numbers or
net Berry fluxes.
In the limit of $L_z \to \infty$, however,
this overlap approaches zero and the Chern number vanishes
in the strictly bulk limit.

Before we close this section, let us discuss about the stability of
the AIAO magnetic ordering in ultra-thin films that may consist of only two or
three alternating layers of Kagome and triangular lattices.
Throughout this section, it is assumed that the AIAO magnetic ordering persists
when the system is prepared in a thin film form.
Though it is reasonable to expect that this is indeed the case when
the film is thick enough, the AIAO may not be stable in the ultra-thin
films due to enhanced spin fluctuations in lower dimensions.
Indeed, a different spin structure with a finite net moment is proposed in recent
theoretical studies of a 1 bilayer system when the contribution of rare-earth
spins is neglected~\cite{Fiete_film2,Fiete_film3}.
In this case, there may exist a Chern insulator phase at intermediate strength of
the interaction~\cite{Fiete_film2,Fiete_film3}.
To investigate the stability of the AIAO magnetic ordering in ultra-thin films including
both the rare earth and Ir spin moments would be
an important issue for future studies.

\subsubsection{Metallic states in domain walls}

The magnetic domain walls are another place where the boundary states
can significantly influence transport and other physical properties.
In the AIAO ground state, all local moments
on one tetrahedron point either inwards to
or outwards from its center, leading to two kinds
of antiferromagnetic phases which are related
by time-reversal symmetry.
Namely, due to the Ising character of the AIAO ordering,
the ground state has two-fold degeneracy, thus
it is natural to expect domain walls between two degenerate ground states.
If some metallic states can be localized at the domain walls,
they can work as additional conducting channels of the system.
Moreover, if metallic surface/interface states exist when
the bulk is fully gapped, it would imply the possibility of controlling the conductivity of the bulk-insulating samples
by varying the number of metallic domain walls with external magnetic fields.

A promising evidence for the existence of domain wall conducting channels is reported by
a recent experiment on Nd$_2$(Ir$_{1-x}$Rh$_{x}$)$_{2}$O$_7$~\cite{Ueda_optical}.
In this system, the substitution of Ir by Rh effectively reduces
the electron correlation as well as the spin-orbit interaction without
changing the electron filling.
The sample with $x=0.1$ shows metallic behavior down to 2K whereas
a full bulk gap of about 45 meV at 5K is opened in the $x=0$ sample.
In the range of $0<x<0.1$, the charge gap is nearly zero
and the optical conductivity appears to be almost linear
with the photon energy, suggesting the possible emergence
of a Weyl SM in the vicinity of the metal-insulator transition~\cite{Hosur_transport}.
In this case, one may expect the existence of conducting channels
coming from the Fermi arc states at the domain boundary.

Through the magneto-resistance measurement and terahertz time-domain spectroscopy,
it is found that the resistivity is much lower in the multi-magnetic-domain state
than in the nearly single-magnetic-domain state.
More specifically, for a given composition $x$, the sample is prepared in
two different ways. One is a trained sample prepared by field-cooling with almost no magnetic domain wall,
and the other is an untrained sample prepared by zero field-cooling, which is expected to have many domain walls.
For the samples with $x=0.02$ and $x=0.05$, which are expected to be Weyl SMs,
the resistivity of the trained sample is much higher than that of the untrained samples,
consistent with the presence of domain wall conducting channels
originating from the Fermi arc states.
Moreover, what is even more interesting is that the domain wall conducting channels
are observed even in the insulating bulk sample with $x=0$.
This strongly suggests that the conducting channels here may have similar origin
as the gapless surface states in the exotic insulating state when $U_{c2}<U<U_{c3}$~\cite{Ueda_DW}.
Furthermore, terahertz time-domain spectroscopy shows
a Drude-like optical response with a small scattering rate about 2 meV,
which is strongly suppressed with increasing field
and irreversibly disappears above 5T.
These observations are consistent with the picture that the magnetic domain wall is highly conductive
with a minimal damping constant of 2meV, contrary to the fully-gapped insulating bulk states.

The presence of metallic states in domain walls, especially when
the bulk states are gapped, is
theoretically suggested by recent unrestricted Hatree-Fock mean field studies by Yamaji and Imada~\cite{Yamaji}.
These metallic states appear when the local spin moments near the domain wall are reduced as compared to the bulk value.
This would decrease the effective $U$ at the domain wall, and hence the effective $U_{c3}$ can be
larger than that of the surface states at the boundary to the vacuum.

Before we close this section, let us note that
there is another, non-topological, source of domain wall metallic states.
Considering that a semi-metal
with a quadratic band crossing point at the BZ center is a candidate non-magnetic ground state,
the vanishing magnetization at the domain wall can naturally have metallic states originating from the bulk semimetal state.
Therefore, distinguishing the topological and non-topological contributions to the domain wall conducting states
would be an important issue for the clarification of the nature of the metallic conduction in multi-domain samples.

\subsection{\label{sec:magnetic_control} Magnetic control of bulk topological properties}

\begin{figure*}[t]
\centering
\includegraphics[width=0.8\hsize]{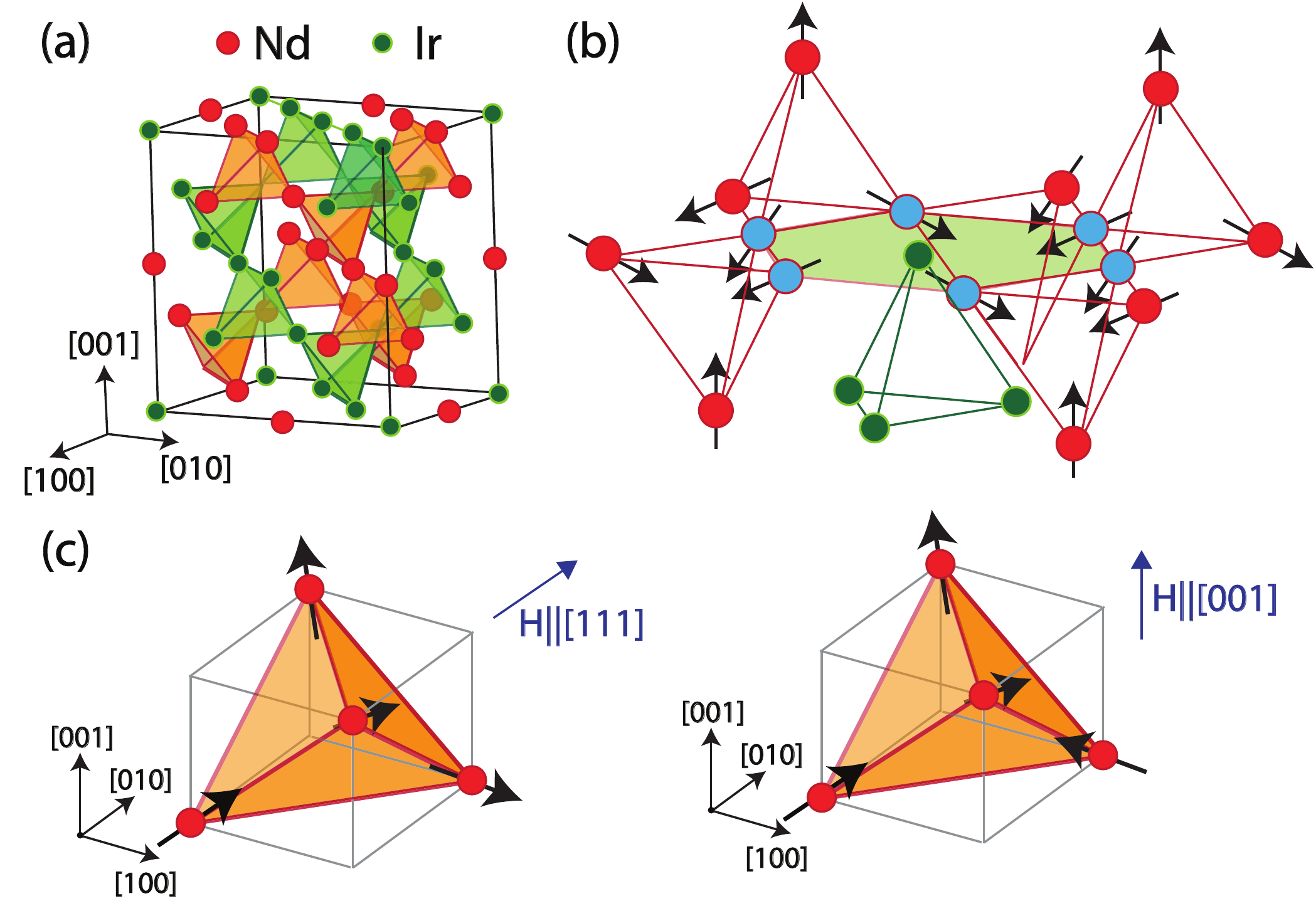}
\caption{
{\bf Lattice structure and f-d exchange couplings in Nd$_2$Ir$_2$O$_7$.}
({\bf a}) Lattice structure of Nd$_2$Ir$_2$O$_7$. Both Nd and Ir sublattices
have their own pyrochlore network.
({\bf b}) Local structure around an Ir tetrahedron.
A single Ir ion is surrounded by six nearest-neighboring Nd sites, marked in blue,
whose local spin moments generate an effective magnetic field at the Ir site at the center.
({\bf c}) Spin configuration of a Nd tetrahedron under the applied magnetic field.
When $B||[111]$ ($B||[001]$), the four spins in a Nd tetrahedron form a 1-in 3-out (2-in 2-out) configuration
due to the Zeeman coupling.
} \label{fig:fdexchange}
\end{figure*}

Up to now, we have considered the bulk and surface properties
of pyrochlore iridates focusing on the role of
Ir 5d electrons only.
However, generally, pyrochlore iridate compounds
with the chemical formula R$_{2}$Ir$_{2}$O$_{7}$,
where R is a rare-earth element,
consist of two interpenetrating pyrochlore lattices occupied by R and Ir, respectively,
which are displaced from each other by a half unit cell.
In many compounds, for example those with R=Nd, Sm, Gd, Dy, Yb, etc.,
both R and Ir may carry local moments.
Owing to the large spin-orbit coupling, R-4f spins possess
a single-ion anisotropy, thus, each spin can be considered
as an Ising spin pointing inward to or outward from the center of each tetrahedron.
On the other hand, in the case of Ir-5d spins, the spin-orbit coupling
(or the Dzyaloshinskii-Moriya interaction in the strong coupling limit)
is the origin of AIAO type spin ordering when the coupling to the rare-earth elements
is neglected. However, once the coupling to the R-4f moments is considered,
the effective magnetic field from the rare-earth moments may also influence the spin anisotropy
in Ir-5d electrons~\cite{Gangchen}.

In particular, the pyrochlore iridate compound with R = Nd offers an ideal platform to
explore the evolution of electronic states with the change
in magnetic configuration where not only the magnetic
moment of Ir-5d electrons but also Nd-4f local moments
play a crucial role.
It is believed that the Ir-5d magnetic moments develop AIAO order below the metal-insulator transition at $T_{\text{MI}}\approx 30K$
in polycrystalline Nd$_{2}$Ir$_{2}$O$_{7}$. Then the effective fields from the Ir-5d moments on the Nd-4f sites arise from the f-d exchange coupling.
This, in turn, induces the Nd-4f moment ordering in the AIAO form. Indeed, recent powder neutron diffraction and inelastic scattering measurements
are consistent with this picture~\cite{Tomiyasu_neutron}. The AIAO ordering of the Ir-5d moments is also confirmed in a related compound
Eu$_2$Ir$_2$O$_7$, where Eu does not carry a local moment at low temperatures~\cite{Nakatsuji_xray}.
From the splitting energy scale of Nd ground-state doublet due to the effective field from Ir moments,
and the dispersion of the relevant excitations, one can extract
the scale of the Nd-Nd exchange energy of $J_{\text{Nd}}\approx 1K$ and
the f-d exchange energy of $J_{\text{fd}}\approx 15 K$~\cite{Tomiyasu_neutron}.
This shows that the Nd ordering pattern
can be strongly modified by the Ir spin configurations through the f-d exchange interaction.

Conversely, we may have a chance of greatly modifying the Ir-5d spin ordering pattern in terms of
magnetic-field control of Nd-4f moments via the
f-d exchange coupling. This will, in turn, change the electronic states on the Ir sites and the corresponding
transport properties.
Note that the external magnetic field will primarily couple to Nd moments
as the magnitude of Ir moments is negligible compared to Nd moments.
Indeed large magneto-resistance is observed in polycrystalline samples, which
indicates that the electronic states are strongly coupled to the magnetic features in this material~\cite{Ueda_DW}.
This idea is further supported by
the recent magneto-transport study with a field-tuned
magnetic structure in a Nd$_{2}$Ir$_{2}$O$_{7}$ single crystal~\cite{Ueda_Bfield,tian2015field}.
In particular, it is found that there exists a large magneto-resistance when the magnetic field $H$
is applied along the $[001]$ direction, while it is much smaller when the field is along $[111]$ direction.

The magnetic-field dependence can be explained by the changes in the magnetic configurations
of the Nd moments, which are sensitive to the magnetic field directions due to small energy
scales of $J_{\text{Nd}}$ and $J_{\text{fd}}$.
For instance, when $H||[111]$ ($H||[001]$),
the magnetic moment configurations on Nd sites
can change from the AIAO pattern at zero field to the 1-in 3-out (2-in 2-out) configuration on
tetrahedra in Nd-sublattice (see Fig.~\ref{fig:fdexchange}) via the Zeeman coupling.
This interpretation is consistent with the values of the saturated-magnetization in
different magnetic-field directions.

Once the Nd moment direction is rearranged, the Ir moment configuration
should also be modified due to the f-d exchange coupling.
Note that the metal-insulator transition temperature in the single crystal
is about $T_{\text{MI}}\approx 15 K$\cite{Ueda_Bfield} or $T_{\text{MI}}\approx 27 K$\cite{tian2015field}, which is comparable to
the f-d exchange scale $J_{\text{fd}}$.
This means the electronic structure of Ir electrons can drastically be changed
by different Nd moment orientations via the f-d exchange coupling.
It has been suggested that this is the fundamental origin of the field direction-dependent
magneto-transport in Nd$_2$Ir$_2$O$_7$ single crystal.

For example, when the magnetic field is applied along [001] direction, the AIAO
magnetic configuration at the Ir sites is fragile to the effective field coming from
the Nd 2-in-2-out magnetic configuration. As a result, the AIAO order at Ir sites and
the associated insulating behavior are suppressed and a semi-metallic state arises,
leading to the large magneto-resistance.
On the other hand, the effective field from Nd 1-in-3-out configuration is less
effective in suppressing the AIAO order on Ir sites and the system remains
largely insulating.

To conclude, the nontrivial interplay of f-d exchange and electron correlation
can induce various intriguing phases in pyrochlore iridates.
In particular, the magnetic field induced switching of Nd-4f moment configurations
may strongly modify the topological nature of the Ir-5d band structure,
which is intimately related to highly anisotropic magneto-transport
properties in Nd$_2$Ir$_2$O$_7$.  Exploration of a family of pyrochlore iridates
with controlling parameters such as $U$ and $J_{\text{fd}}$ in terms of the substitution
of R ions and the application of magnetic fields in different directions may reveal
the emergence of abundant exotic phases.

\section{\label{sec:3d}Three-dimensional honeycomb iridates}

In this section, we will examine the recent theoretical and experimental progress on the tri-coordinated iridate materials $A_2$IrO$_3$ and the associated models, with a focus on the three-dimensional \liiro{} compounds. These systems exemplify the physics of spin-orbit coupling and have drawn considerable recent interest as a result. In particular, the anisotropic nature of the spin exchange interactions is critical to the physics in these systems; as we will see, this gives rise to fascinating phases of matter which depend on the strong SOC for their presence.

Bond-anisotropic models have been continually studied in literature even before the recent discoveries of iridate materials. The general class of models known as compass models are an example of such, where the pseudospins interact in a directionally-dependent manner and can represent degrees of freedom like orbitals, vacancy centers, or even atomic states in cold atom systems.\cite{nussinov2015compass} 
Similar to these compass models is the Kitaev model on the honeycomb lattice,\cite{kitaev2006anyons} which is of particular significance to the iridate materials. This anisotropic spin-$1/2$ model can be exactly solved to reveal a spin-liquid ground state that can be used to perform fault-tolerant topological quantum computation.\cite{Kitaev2003} The relation of the Kitaev model to iridate materials was first elucidated by the seminal work of Jackeli and Khaliullin\cite{jackeli2009mott}: using a perturbative calculation in the strong coupling limit, these authors invigorated the field by conjecturing that the honeycomb iridate \liiro{} can potentially realize the Kitaev model and its spin liquid ground state. However, both \liiro{} and its isoelectronic cousin \nairo{} have been found to order magnetically\cite{Ye2012dr,choi2012spin,liu2011long,gretarsson2013magnetic} and subsequently many studies have sought an explanation for the presence of magnetic order in place of a quantum spin-liquid ground state.\cite{mazin2012na,yamaji2014first}

Crystal distortions and further neighbour interactions in these quasi two-dimensional (2D) honeycomb iridates have been implicated as two possible causes for the observed magnetic ordering.\cite{rau2014trigonal,kimchi2011kitaev,katukuri2014kitaev,yamaji2014honeycomb,sizyuk2014importance} If true, the Kitaev spin liquid ground state may be difficult to realize in these materials since a controlled elimination of these effects is beyond the reach of current experimental techniques. It was, therefore, a welcomed development when two three-dimensional honeycomb-like iridates were independently and simultaneously discovered in 2013.\cite{Modic2014ch,takayama2015hyperhoneycomb} These two materials, the hyperhoneycomb iridate \bliiro{} and the stripyhoneycomb iridate \gliiro{}, are polymorphs of the quasi 2D \airo{}. Not only are they theoretically compatible with the Kitaev model and an extension of its exact spin-liquid ground state solution, their fully three-dimensional lattice structure induces physics that is distinct from their two-dimensional counterparts. In addition, the limited experimental viability of studying iridium-based materials using neutron scattering\cite{choi2012spin} have pushed researchers to refine resonant x-ray scattering probes,\cite{gretarsson2013magnetic,chun2015direct} which was used to characterize the ground states of these 3D iridates in unparalleled detail.

Although these two 3D honeycomb iridates also magnetically order, and are thus not examples of a quantum spin liquid, the study of their behaviour offers another window through which we can capture another facet of spin-orbit coupling effects in correlated materials. In the rest of this section, we will survey recent works on \bliiro{} and \gliiro{} to demonstrate the rich physics present in this area of study. We set the stage of discussion by first describing the experimental signatures of \bliiro{} and \gliiro{}, which includes their crystal structures, thermodynamic properties, and magnetic orderings. Next, we construct a microscopic understanding of iridates by examining the nature of their spin-orbital entangled wavefunctions, which will enable us to investigate the anisotropic interactions between them. We will then explore the Kitaev model on these 3D lattices with an emphasis of how they differ from their 2D counterpart. Lastly, we will delve into some recent theoretical proposals for the observed magnetic ordering and their relation to the Kitaev spin-liquid.

\subsection{\label{sec:3d_experiment}Crystal structure and experimental signatures}
\begin{figure}[h!]
  \centering
  \setlength\fboxsep{0pt}
  \setlength\fboxrule{0pt}
  \subfloat[][Hyperhoneycomb lattice]{
    \label{fig:h0_lattice}
    \fbox{\includegraphics[scale=.11,clip=true,trim=400 0 400 0]{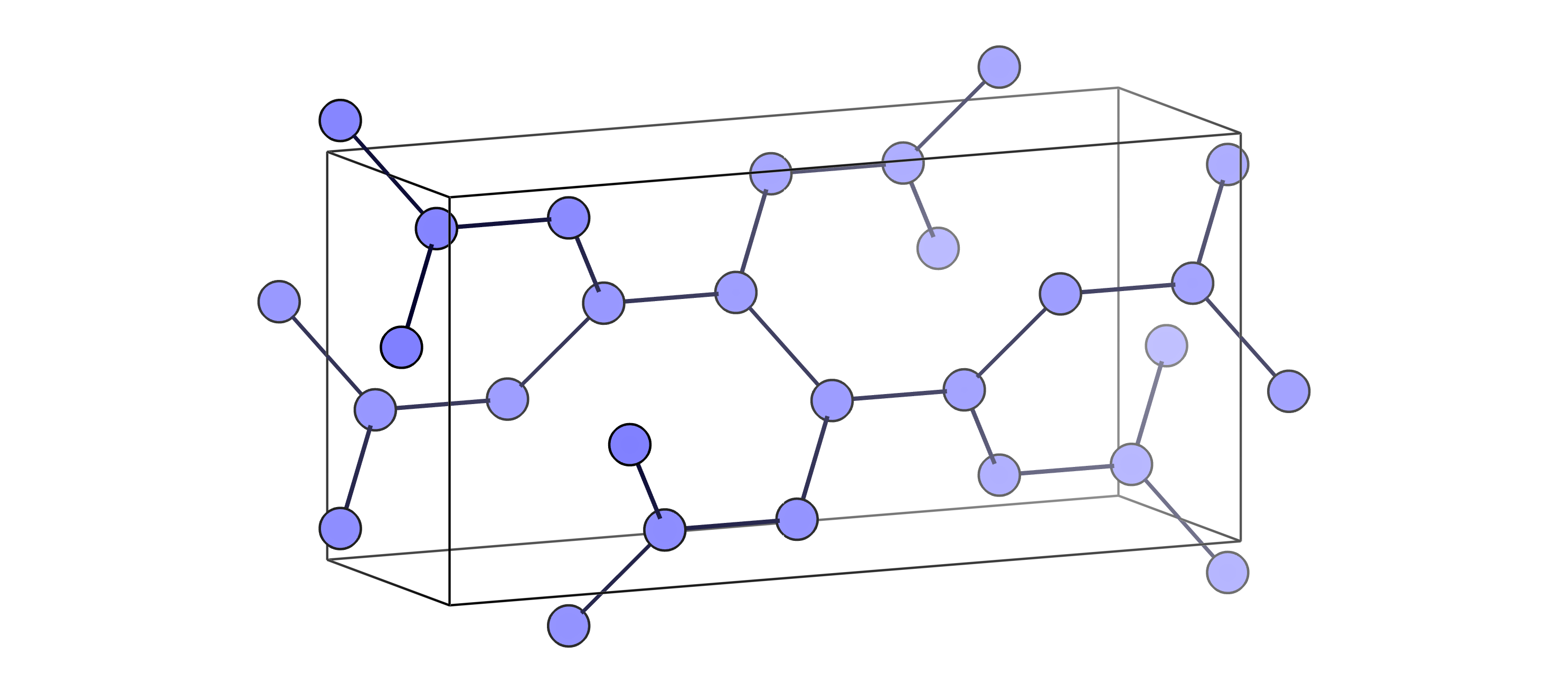}}
  }

  \subfloat[][Stripyhoneycomb lattice]{
    \label{fig:h1_lattice}
    \fbox{\includegraphics[scale=.12,clip=true,trim=500 0 500 0]{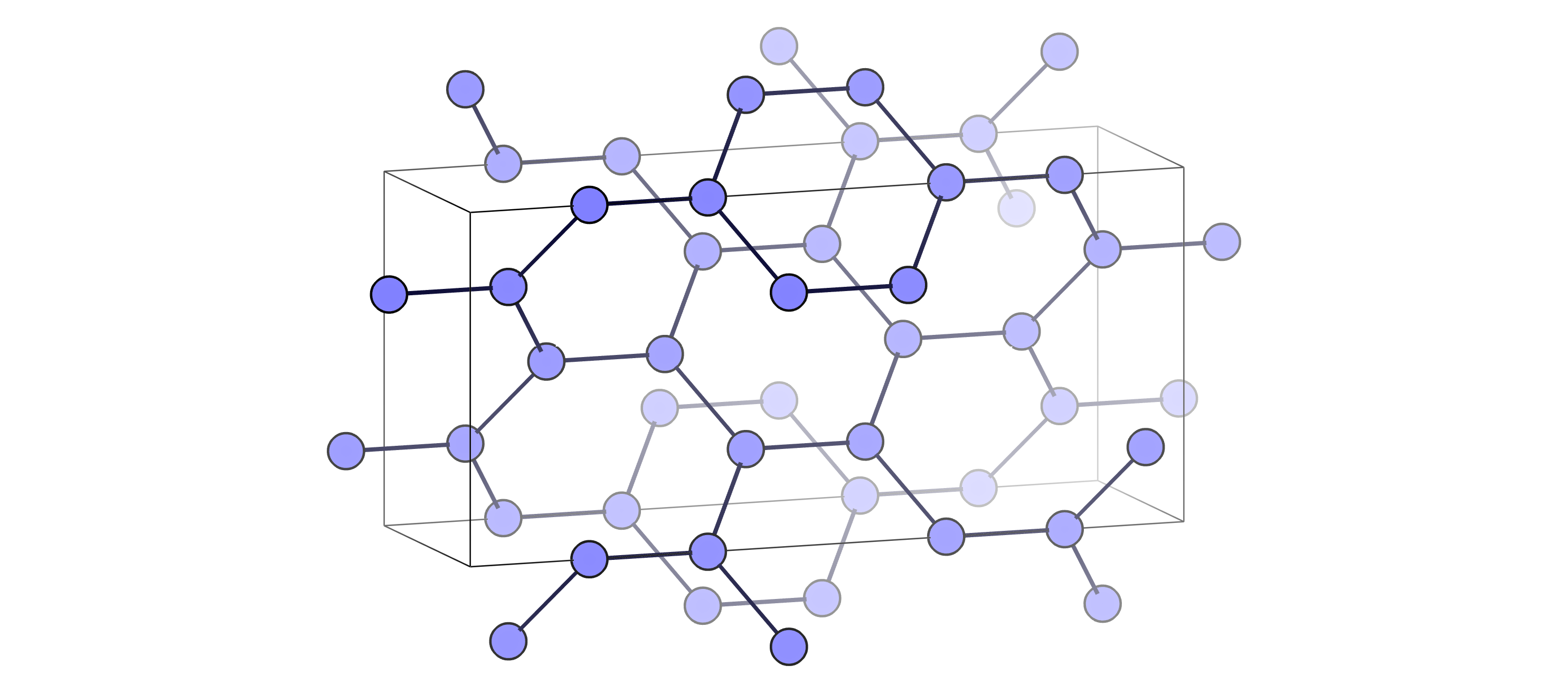}}
  }
  \caption{\label{fig:lattices}
  {\bf Iridium ion placements in the (a) \harzero{} and (b) \har{} lattices.} The primitive unit cells have four and eight atoms, and are in an face centred and base centred structures, respectively. Here, the conventional unit cell is portrayed.
}
\end{figure}

The hyperhoneycomb \bliiro{} and stripyhoneycomb \gliiro{} are both fully three-dimensional crystal structures as shown in Fig. \ref{fig:lattices}.  Determined by single crystal X-ray analysis, both materials are orthorhombic: the former is in the face-centred space group $Fddd$ (No. 70), the latter is in the base-centred space group $Cmmm$ (No. 66). They have similar lattice constants, and both contain the same number of atoms in their conventional unit cells. The basic building block present in both structures is the IrO$_6$ octahedron: these octahedra are \textit{tri-coordinated} (each have three nearest-neighbours) and adjacent octahedra are \textit{edge-shared} (they have two oxygen atoms in common). Depending on how these building blocks are arranged in the crystal, a variety of structures---including the 2D honeycomb iridates---can be constructed.

To describe the arrangement of octahedra in these various structures, we first note that there are several ways to orient a tri-coordinated and edge-shared cluster of four octahedra even if we stipulate that the octahedral axes are aligned with a set of global axes.\footnote{In this construction, we aim to understand the connectivity of the iridium lattice, hence we neglect the distortions that are present in the real crystals.} In the 2D honeycomb iridates, the orientation of these clusters ensure that all iridium sites are coplanar, resulting in a two-dimensional layer that forms the real crystal structure when stacked. In the 3D honeycomb iridates, there exists six-octahedra clusters where the middle bond twists, as demonstrated in Fig. \ref{fig:lattices}. These twists only occur along bonds that are parallel to the orthorhombic $c$ direction (hereafter we call these the $z$-bonds), hence both 3D structures possess zigzag chains similar to the 2D honeycomb lattice. The difference between the two 3D lattices is the frequency of the bond twists: in \bliiro{}, they happen at every $z$-bond, while in \gliiro{}, these twists happen at every other $z$-bond. The resulting hyperhoneycomb lattice does not possess any hexagons and the smallest loop of iridium sites contains ten ions. On the other hand, the stripyhoneycomb lattice does possess strips of hexagons. This construction of the crystal structures enable the generalization of a family of polymorphs of \liiro{} called the \textit{harmonic honeycomb series}, denoted as $\langle n \rangle$-\liiro{}, where $n$ indicates the number of bonds between adjacent bond twists. In this nomenclature, \bliiro{}, \gliiro{}, and \aliiro{} are $\langle 0 \rangle$-, $\langle 1 \rangle$-, $\langle \infty \rangle$-\liiro{} respectively.\cite{Modic2014ch}

Although the Li$_2$IrO$_3$ compounds can be represented by iridium atoms lying in the center of oxygen octahedra in an ideal structure, deviations from this simple model are present. In particular, deviations in the bond length between the different bond types in the model exist in the real crystals, along with deviations from the ideal bond angle between the different atoms. These deviations are smaller in the three-dimensional compounds \bliiro{} and \gliiro{} than in the two-dimensional \aliiro, but may still be relevant to the physics of these compounds. As a result of these, additional interactions which are forbidden by symmetry in the simple model can be present in the real materials, with strength proportional to the deviation from the ideal. While we do not discuss this in detail in this review, this can have impotant effects on the magnetism which appears in these models.

\begin{figure}[h!]
  \centering
  \setlength\fboxsep{0pt}
  \setlength\fboxrule{0pt}
  \subfloat[][Spin ordering on the hyperhoneycomb lattice]{
    \label{fig:h0_spiral}
    \fbox{\includegraphics[scale=.045,clip=true,trim=2000 0 1800 0]{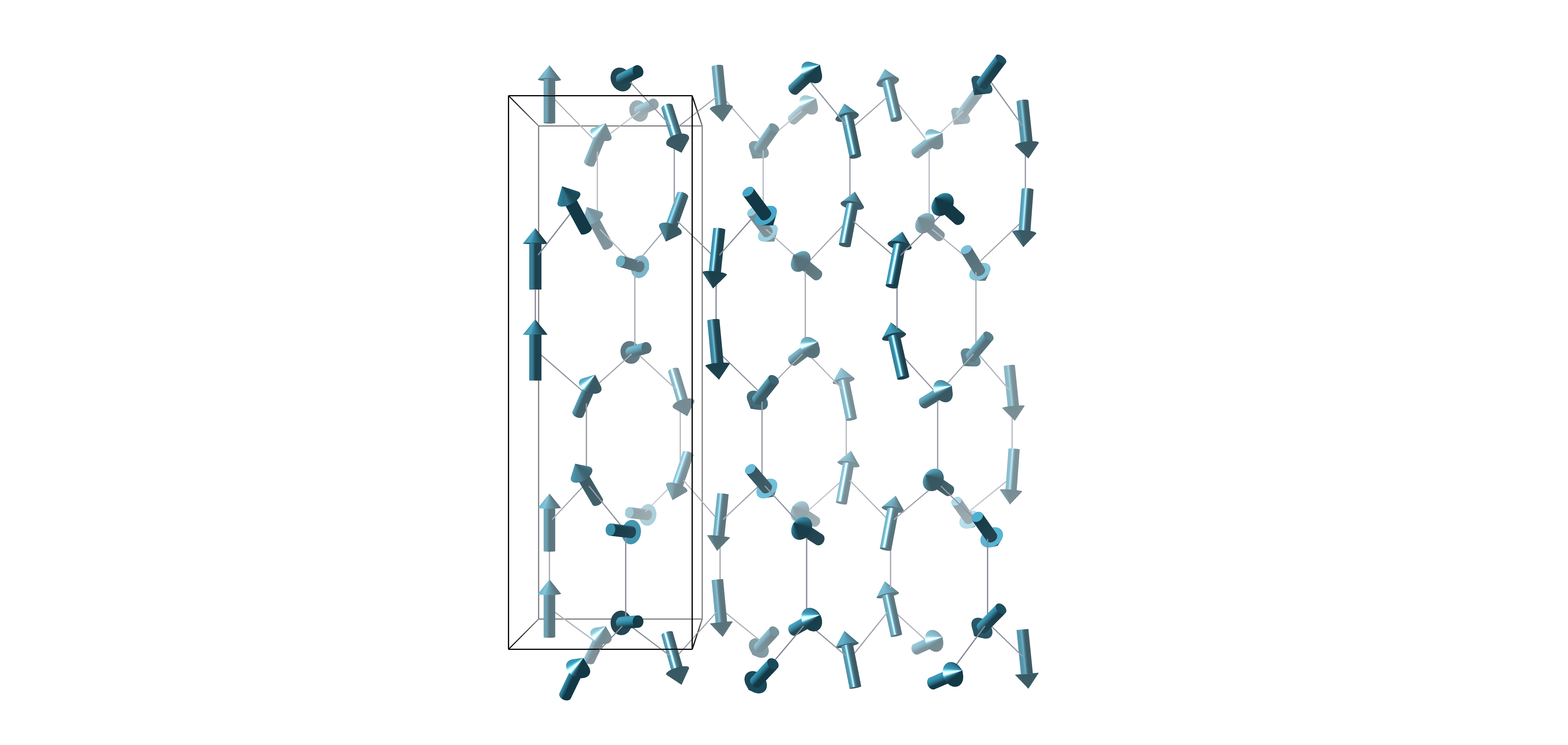}}
  }

  \subfloat[][Spin ordering on the stripyhoneycomb lattice]{
    \label{fig:h1_spiral}
    \fbox{\includegraphics[scale=.042,clip=true,trim=1800 0 1800 0]{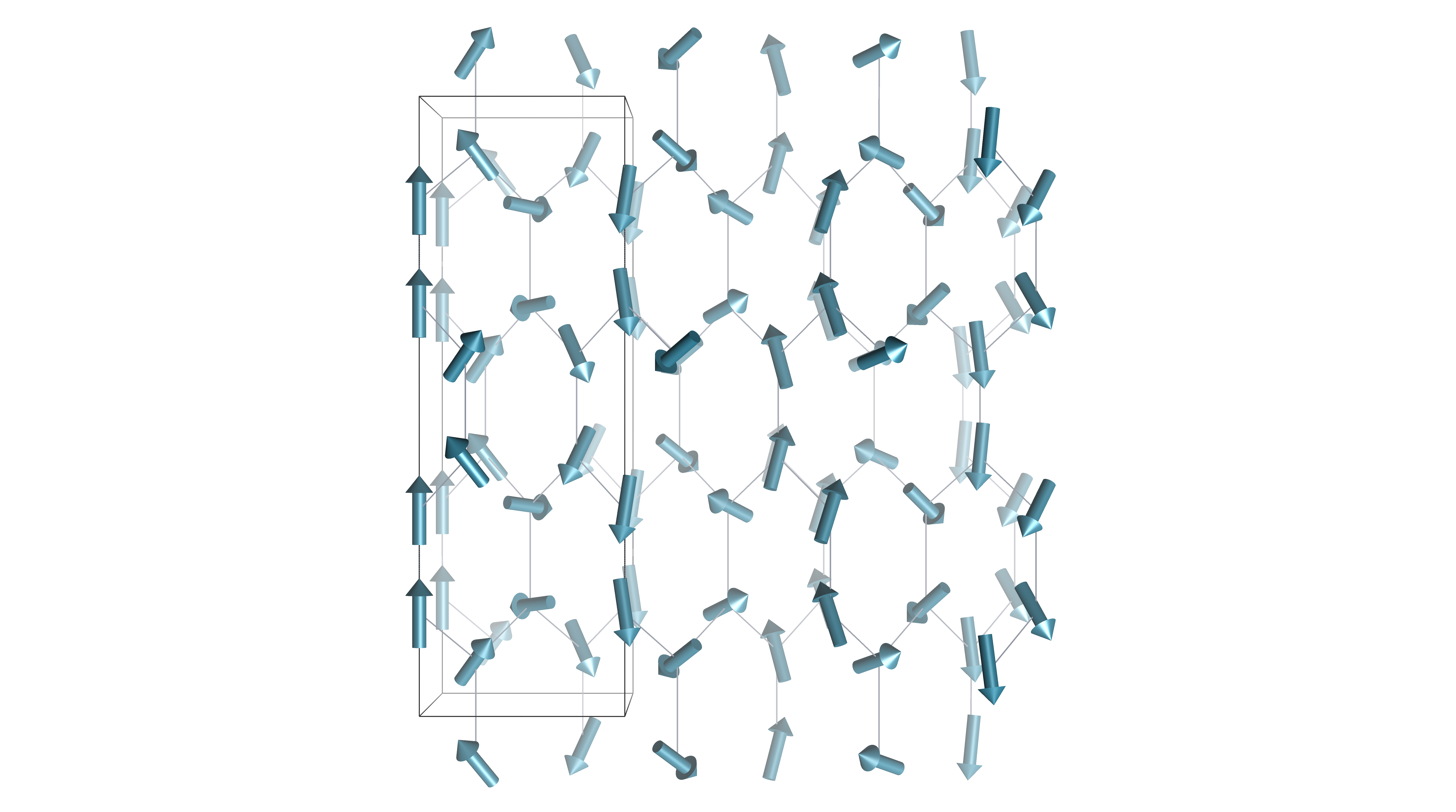}}
  }
  \caption{\label{fig:lattice_phases}
  {\bf The experimental magnetic moments of the spiral phase observed
in the (a) \harzero{} and (b) \har{} lattices.} These were determined in Ref. \onlinecite{biffenbet} and \onlinecite{biffengam} respectively. The propagation
wavevector is in the a direction and the moments on adjacent sublattices
counter rotate as they propagate.
}
\end{figure}

Not only do the two 3D honeycomb iridates have similar structures, they also possess remarkably similar experimental signatures. Magnetic susceptibility measurements indicate both materials follow Curie-Weiss behaviour with effective moment sizes close to that of a spin-$1/2$ degree of freedom.\cite{Modic2014ch, takayama2015hyperhoneycomb} Both iridates undergo antiferromagnetic transitions at $\sim 38~K$ yet the Curie-Weiss temperatures suggest dominant ferromagnetic exchange interactions. Using a combination of neutron and magnetic resonant X-ray scattering, the magnetic orders of both materials were revealed to be non-coplanar, incommensurate spiral phases where moments on neighbouring sublattices rotate in the opposite sense (also known as \textit{counter-rotation} of spirals), as seen in Fig. \ref{fig:lattice_phases}. Remarkably, the ordering wavevectors in both materials coincide within experimental uncertainty and are given by $\vec{q}\sim[0.57,0,0]$.\cite{biffin2014unconventional, biffin2014non}

Additional probes were used to investigate the magnetism in both materials. Upon cooling \gliiro{} in the high temperature paramagnetic phase, torque measurements revealed a reordering of the principal magnetic axes, which was attributed to the presence of spin-anisotropic exchanges.\cite{Modic2014ch} In \bliiro{}, the application of a small external magnetic field ($\sim3~T$) was able to drive the system into a ferromagnetic state. The orbital contribution to the induced magnetic moment was investigated by X-ray magnetic circular dichroism, which suggested that the ratio of spin to orbital moment is closed to that of the ideal \jhalf{} moment. External pressure was also applied to \bliiro{} in a magnetic field. The insulating behaviour of \bliiro{} persisted above 2 GPa, but the field-induced ferromagnetic moments was quickly suppressed. This was interpreted as the presence of other almost degenerate states near the magnetic ground state.

The unconventional magnetism observed in both materials together with their striking experimental similarities prompted theoretical studies on the ground states of these materials and their relation to the Kitaev spin liquid. As both materials exhibit Mott insulating behaviour, most of these theoretical works approached the problem from the strong correlation limit. In the next section, we examine several important results that provided the microscopic basis for some of the models used in literature in understanding the behaviour of these 3D honeycomb iridates in the strong correlation limit.

\subsection{\label{sec:3d_exchange}Strong correlation limit and spin-orbit coupling}

In terms of the \ttg{} states, the \jhalf{} doublets which describe the low energy subspace in the presence of strong correlations and SOC can be written as
\begin{equation}
	\label{eq:jeff}
	\ket{\frac{1}{2},\pm\frac{1}{2}} =\sqrt{\frac{1}{3}}\left(
	\ket{yz,\mp} \pm i\ket{xz,\mp}
	\pm \ket{xy,\pm}
	\right),
\end{equation}
where $\ket{yz,\pm},\ket{xz,\pm},\ket{xy,\pm}$ are the \ttg{} states and $\pm$ correspond to spin up and down. As we see, the presence of large \soc{} entangles the spin and orbital degrees of freedom in these localized wavefunctions.
Despite the seemingly anisotropic spatial and spin dependence of these wavefunctions, the projected magnetic moment operator in this subspace is proportional to the $J$ operator with an isotropic $g$-factor of $-2$.

The \jhalf{} pseudospins which appear in the presence of strong spin orbit coupling and strong electronic correlations are composed of spin-orbital entangled wavefunctions that transform like a pseudovector under the symmetries of the octahedral group.  Using this knowledge, the symmetry-allowed pseudospin model can be worked out explicitly by considering the crystal symmetries of the lattice.  Perturbative calculations\cite{jackeli2009mott,chaloupka2010kitaev,rau2014generic} have motivated the study and understanding of highly anisotropic exchange parameters, while first principles calculations\cite{foyevtsova2013ab,kim2015predominance} have verified the validity of the \jhalf{} basis in the case of \bliiro{}.

Even in an ideal lattice (in the absense of distortions), a large number of spin exchange parameters are allowed by the crystal symmetries of the 3D lattices. Considering only nearest neighbour spin interactions, 10 terms are allowed by symmetry for the \harzero{} lattice, with more allowed for higher harmonic lattices. However, if we consider the model which arises from a strong coupling expansion of the Kanamori Hamiltonian, only 3 free parameters remain; a standard Heisenberg coupling, a symmetric off-diagonal exchange term and the bond-anisotropic Kitaev term. It is this final term that has driven much of the interest in these systems, and the model with this term alone has been the focus of much research.

\subsection{\label{sec:3d_kitaev}Kitaev spin liquid}

When only the Ir-O-Ir exchange process is present between Ir atoms in an ideal edge-shared octahedral environment, the spin model which appears at strong coupling is the highly anisotropic Kitaev Hamiltonian.\cite{jackeli2009mott}
The realization that a model with such possible interest to experimentalists and theorists alike could arise in a physical system with strong spin-orbit coupling led to a flurry of research into these systems.\cite{chaloupka2010kitaev} In order to search for the Kitaev spin liquid in experiments, we must first understand the theoretical predictions of the model.

The Kitaev model was first proposed on a 2-d honeycomb lattice by Alexei Kitaev,\cite{kitaev2006anyons} as an example of a nearest neighbour spin model which hosts a spin-liquid ground state, with excitations which could be used to perform quantum computation.\cite{Kitaev2003} The Hamiltonian of this model takes the form
\begin{eqnarray}
  H_K = \sum_{\langle ij \rangle \in \alpha} K_{\alpha} S_i^{\alpha}S_j^{\alpha},
\end{eqnarray}
where $\alpha$={x,y,z} denote the three bond directions on the honeycomb lattice as well as the three components of the S=1/2 spin operators, and the sum runs over nearest neighbour bonds \{$ij$\}. This system can be exactly solved by replacing the spin operators on each site with four Majorana fermions \{$b^x,b^y,b^z,c$\} using $S_i^\alpha = ib_i^\alpha c_i$. The operators defined as $u_{ij} = ib_i^\alpha b_j^\alpha$ on an $\alpha$ bond commute with one another and with the Hamiltonian, and are therefore constants of motion. We can block diagonalize the Hamiltonian into sectors in which the operator $u_{ij}$ is replaced by its eigenvalues, $\pm 1$. However, we have expanded the Hilbert space by replacing the spin operators with Majorana fermions; as a result, the eigenvalues of $u_{ij}$ are not gauge invariant. The gauge invariant quantities can be shown to be the products of the $u_{ij}$ operators around loops, which determine the fluxes of the $Z_2$ gauge theory in the model. The ground state of this system must have zero flux passing through each plaquette, due to a mathematical theorem known as Lieb's theorem.\cite{lieb1994flux} Choosing $u_{ij}$ in a symmetric fashion which satisfies this constraint results in a quadratic Hamiltonian for the $c$ fermions. This Hamiltonian can be diagonalized; the resulting spectrum depends of the relative values of $K_\alpha$. If $|K_\alpha| \leq |K_\beta| + |K_\gamma|$ for all choices of $\alpha, \beta, \gamma \in \{x,y,z\}$, the spectrum is gapless, with the nodes appearing as two Dirac points if this inequality is strict. If this is not the case, the spectrum is gapped.

The exact solution to the Kitaev model on the Honeycomb lattice relied on the large number of conserved quantities in the model (the $u_{ij}$ operators). However, any lattice in which each site is connected to three others by bonds of the $x,y$ and $z$ types can be solved using the same procedure. In particular, the \harzero{} and \har{} lattices of iridium atoms found in \bliiro{} and \gliiro{} satisfy these conditions. This offers an avenue to explore spin liquid physics in three dimensions, which arises in realistic models from strong spin-orbit coupling.\cite{Mandal2009,Modic2014ch}

While the three-dimensional Kitaev model is similar to the two-dimensional version, a number of important differences arise. On the infinite honeycomb lattice the fluxes on each plaquette can be chosen to be zero or $\pi$ independently, whereas on the three-dimensional lattices the fluxes are forced to obey a constraint; the sum of fluxes of a set of loops enclosing a volume without holes must be equal to zero (modulo 2$\pi$).\cite{Mandal2009} As a result, any non-zero fluxes in the system must appear in loops. In the honeycomb lattice Kitaev model, the state with finite non-zero flux density continuously connects with the high-temperature phase, meaning a finite temperature phase transition is impossible. In contrast, in the three-dimensional models the spin liquid state can persist to finite temperatures, with a transition to the high-temperature phase occurring when the length of the loops diverges.\cite{senthil2000z} This has been studied numerically using monte carlo simulations, and the finite temperature properties of this phase has been extracted.\cite{Nasu2014ft,nasu2014vaporization}

In addition, due to the lack of reflection symmetry of the three-dimensional lattices, Lieb's theorem no longer applies, and therefore there is no reason to expect a zero-flux ground state. On the \harzero{} lattice, numerical studies suggest that the ground state has zero flux,\cite{Mandal2009} however on the \har{} lattice $\pi$ flux must pass through a subset of the loops.\cite{Schaffer2014ts}

Once the ground state flux sector is identified, we can examine the spectrum of excitations of Majorana fermions. Of particular interest are the low energy excitations, specifically the presence and properties of zero energy Majorana fermion excitations. On the \harzero{} and \har{} lattices, these gapless points appear as a nodal ring in momentum space, in contrast to the zero-dimensional Dirac points which appear on the honeycomb lattice.\cite{Mandal2009,Modic2014ch,Schaffer2014ts,mullen2015} Similar to the honeycomb Kitaev model, if the anisotropy between the couplings on different bonds grows beyond a critical value, a gap opens in the spectrum.

\begin{figure}[h!]
\subfloat[][Nodal Ring]{
    \label{fig:nodal}
    \includegraphics[scale=.55,clip=true,trim=20 20 20 20]{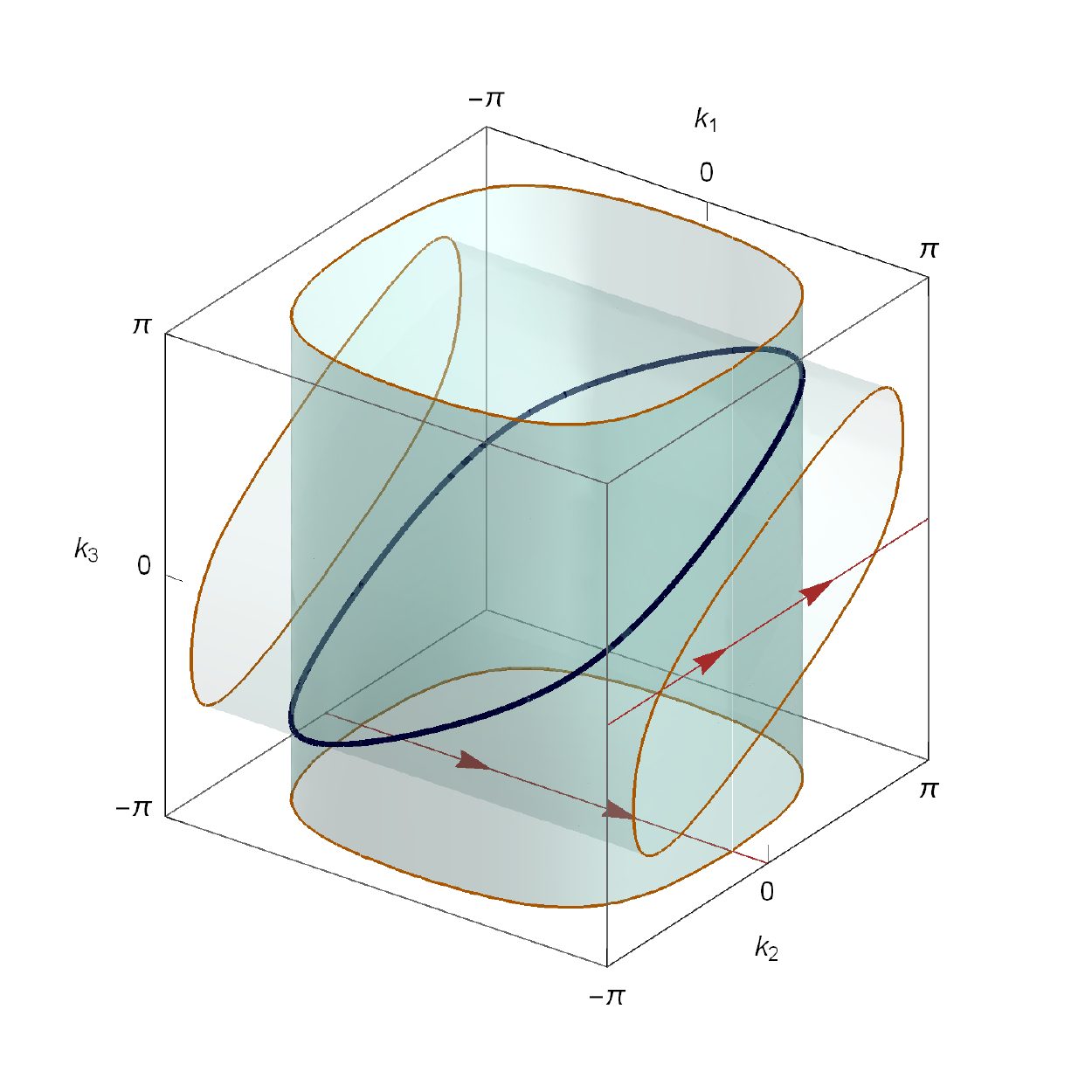}
  }
{
\begin{minipage}[h!]{0.16\textwidth}
\includegraphics[width=\linewidth,keepaspectratio=true]{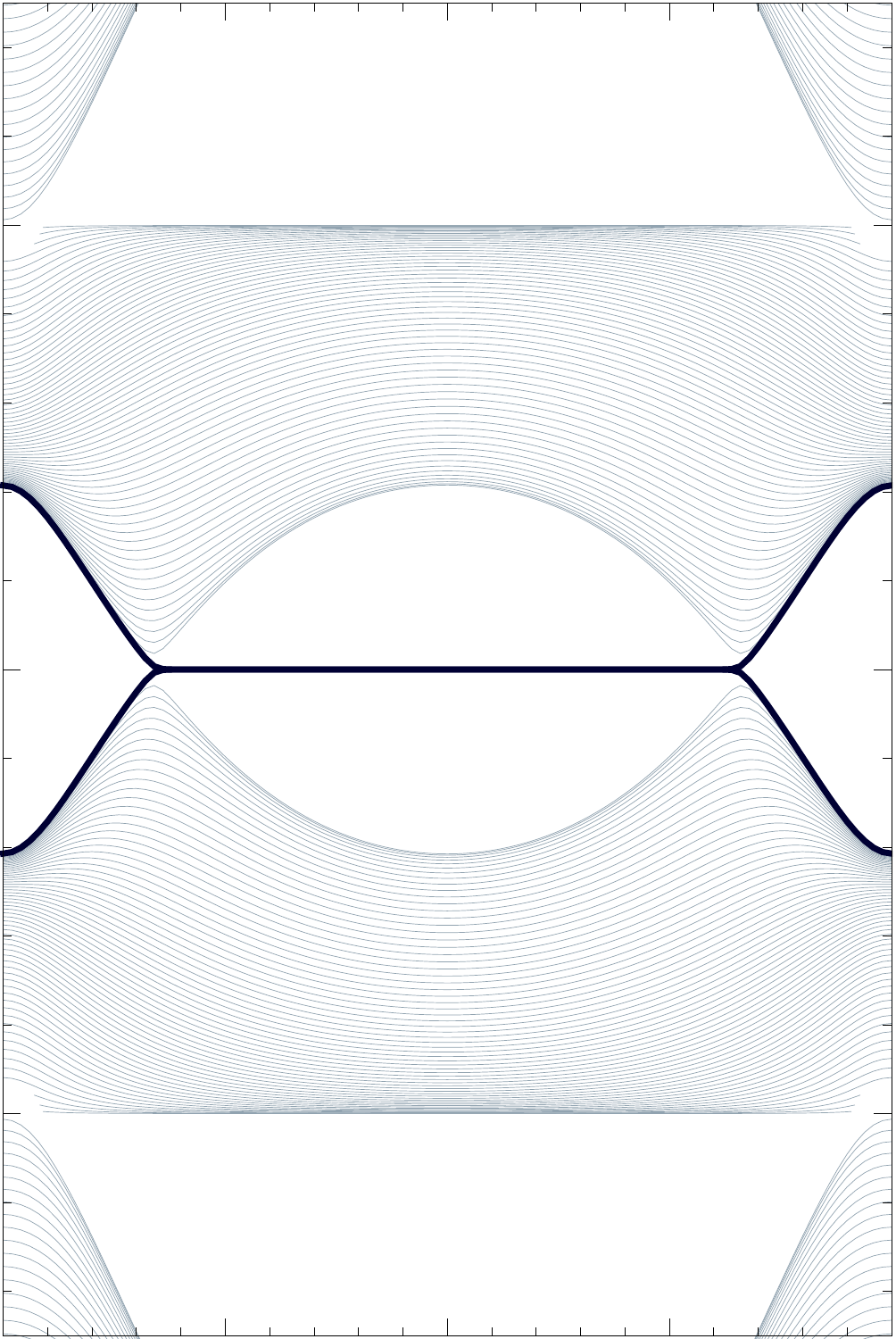}
(b) Front Surface
\label{fase1}
\end{minipage}
\hspace*{.8cm} 
\begin{minipage}[h!]{0.16\textwidth}
\includegraphics[width=\linewidth,keepaspectratio=true]{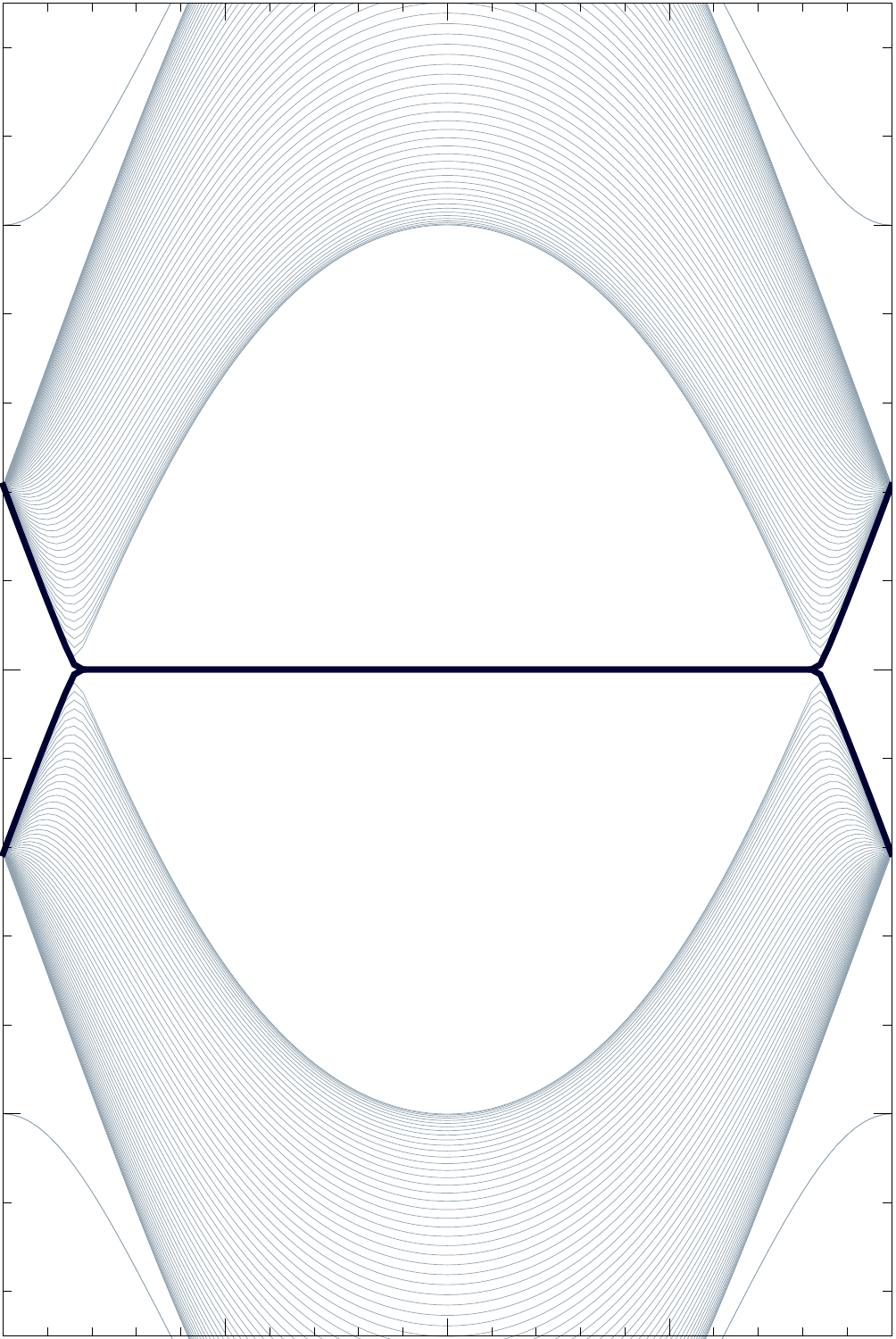}
(c) Bottom Surface
\label{fase2}
\end{minipage}
}
\caption{{\bf The Majorana fermion excitation spectrum of the Kitaev model on the \harzero{} lattice.} ({\bf a}) The location of the nodal ring in momentum space, where
k1, k2, and k3 are coordinates of the primitive reciprocal lattice
vectors. The projection of the nodal ring to the front and bottom surfaces of the
Brillouin zone are shown in orange. ({\bf b}) The spectrum of Majorana fermions which appears on the front surfaces, with flat bands appearing due to the bulk-boundary correspondence. Here, the cut is along the line indicated in red on figure (a). ({\bf c}) Similar to (b), on the bottom surface.}
\end{figure}

The Kitaev spin liquid has been identified as the ground state of the Kitaev model on the \harzero{} and \har{} lattices. However, for this state to appear in physical materials it must also be stable against small perturbative interactions which may be present. Fortunately, this is the case, as an RG analysis shows that short-range four fermion interactions in the Majorana Hamiltonian are irrelevant.\cite{Lee2014} These terms correspond to two spin terms in the Kitaev Hamiltonian; thus, the spin liquid phase is stable with respect to small, short-range spin interactions.

It is worthwhile to also examine the stability of the nodal ring which appears in this spectrum. In the Majorana fermion Hamiltonian both time-reversal and particle-hole symmetry are present, with eigenvalues that imply that the system is in the symmetry class BDI.\cite{Schnyder2008,Schaffer2014ts,hermanns2014quantum} As a result, the nodal ring is characterized by an integer valued topological invariant, associated with the winding number around the nodal ring. This ring is therefore protected, as long as the symmetry remains unbroken.

In addition to indicating the topological stability of the nodal ring, the presence of a non-zero topological invariant in this system implies the existence of zero-energy flat bands in the surface spectrum.\cite{matsuura2013protected,Schaffer2014ts} These bands appear on surfaces with finite projection of the nodal ring, as can be confirmed by direct computation. These flat bands on the surface offer a possible route to identify the Kitaev spin liquid in real materials. By examining the thermal transport properties along the different surfaces in this model, the presence of these flat bands can be identified. In addition, at sufficiently low temperatures, the dominant contribution to the specific heat in such a system would be due to the surface modes.

In the absense of time reveral symmetry, the nodal ring is no longer protected. In the presense of a perturbative magnetic field the nodal ring spectrum on the \harzero{} lattice gains a gap, except at two topologically protected Weyl points of opposite chirality. The flat surface bands which were present also gain a gap, with a gapless Fermi arc connecting the projections of the Weyl points on the surface. With a further increase in the strength of the time reversal breaking term, these Weyl points each split into three - two with the chirality of the original point, and one with the opposite chirality. Therefore, these are all topologically protected objects in this system.\cite{hermanns2015weyl}

We have primarily focussed on the \harzero{} and \har{} lattices, due to the experimental discovery of these lattices in $\beta$- and \gliiro. However, the Kitaev model can be defined on other three-dimensional lattices as well, and can be solved exactly if the lattice is tri-coordinated. In particular, on any of the harmonic honeycomb series of lattices, the Kitaev model can be studied.\cite{Modic2014ch,Schaffer2014ts} In these cases, the results appear to be similar to those on the \har{} lattice, with a non-zero flux sector being the ground state and a topologically protected nodal ring spectrum appearing. The Kitaev model has also been explored on the hyperoctagon lattice; in this case, the zero energy modes form a two-dimensional fermi surface.\cite{hermanns2014quantum}

Having explored many theoretical aspects of the Kitaev spin liquid, it is important to also consider how one would identify its presense experimentally. The two primary methods which have been suggested for doing so are Raman spectroscopy\cite{knolle2014raman,Perreault2015} and inelastic neutron scattering.\cite{knolle2014dynamics,Smith2015neutron} The Raman response of this state has been explored using the Loudon and Fleury approach, which predicts a highly anisotropic response spectrum with a peak structure which could be identified in experiments for both the \harzero{} and \har{} lattices. Inelastic neutron scattering measures the dynamical spin structure factor of the system, which can be directly computed in this state. Two distinctive features emerge in the Kitaev spin liquid state: an energy gap under which no scattering occurs, and diffuse response at high energies. The energy gap appears due to the fact that single spin excitations must cause an excitation in the flux sector, which requires a finite amount of energy. Diffuse scattering is a more general property of spin liquids, and indicates the presense of fractionalized excitations.

\subsection{\label{sec:3d_magnetism}Magnetism}

While the Kitaev model is of interest on its own, without additional interactions it cannot explain the magnetism which appears in experiments on \bliiro and \gliiro. The exact solution of this model finds a spin liquid ground state, while the experiments find a spiral magnetic order. Therefore, additional terms in the Hamiltonian must be considered, if we wish to explain the experiments. While many possible terms exist, it is natural to consider the remaining terms which appear in the strong coupling expansion first. 

In the case of the two-dimensional honeycomb structure, the Heisenberg-Kitaev (HK) model was initially suggested as a minimal model for describing the magnetism present in \nairo.\cite{chaloupka2010kitaev} This model, in which an antiferromagnetic nearest neighbour Heisenberg interaction is present in addition to a ferromagnetic Kitaev interaction, can be shown to have three phases. Firstly, a Neel antiferromagnetic phase is found when the Heisenberg term is dominant. Secondly, a spin liquid phase, with the Kitaev model being a special point, is found when the Kitaev term is dominant. Finally, a stripy antiferromagnetic phase, in which stripes of ferromagnetically aligned spins are connected antiferromagnetically with one another, is present when the Heisenberg and Kitaev terms have similar strength. An extension of this model, in which the signs of the terms are allowed to vary, was also considered; in this case, a ferromagnetic phase and stripy antiferromagnetic phase are also present.\cite{chaloupka2013zigzag}

With the discovery of \bliiro, the HK model was first explored for the three-dimensional hyperhoneycomb (\harzero) lattice, as a possible model of the magnetic order present.\cite{Lee2014} A four-sublattice rotation, first considered on the 2-d honeycomb structure, is present on the \harzero{} lattice as well; using this, and the exact solution of the Kitaev model, we can positively identify three dimensional analogs of each of the orders present on the honeycomb lattice. As discussed in the previous section, the spin liquid phase is stable with respect to nearest neighbour two-spin interactions; as a result, we expect each of these phases to be present over a finite portion of the phase diagram. Classical Monte-Carlo simulations suggest that these are the only magnetic phases present in this model. Similar conclusions are also valid for the \har{} lattice system.

While the ground states of the HK model on the \harzero{} lattice resemble those found on the honeycomb lattice, the finite temperature properties differ significantly. In two dimensions, the Kitaev spin liquid and the SU(2) (spin) symmetric points in each magnetic phase are known to be unstable at any finite temperature. In contrast, in three dimensions, each of these phases is stable at finite temperature, with an ordering temperature estimated to be on the order of three-fifths of the Heisenberg coupling strength for much of the phase diagram.\cite{lee2014order} From finite temperature and quantum order-by-disorder calculations, a particular direction in spin space is chosen from the classically degenerate manifold away from the SU(2) symmetric points; thus, only a finite set of expected spin polarisations are favoured.

Although the HK model is of theoretical interest in these systems, all of the predicted magnetically ordered ground states have a $q=0$ ordering vector. This contrasts with the spiral ordering vector seen in resonant x-ray diffraction and neutron scattering measurements, which indicate a wavevector of [0.57,0,0] for the magnetic order.\cite{biffin2014unconventional, biffin2014non} As such, additional interactions are required in order to capture the magnetic order in these systems.

In the strong coupling expansion of the Kanamori Hamiltonian for both the 2-d and 3-d honeycomb lattice systems, an symmetric off diagonal exchange term appears in addition to the Heisenberg and Kitaev terms.\cite{rau2014generic} The resulting Hamiltonian takes the form
\begin{eqnarray*}
  H = \sum_{\langle ij \rangle \in \alpha \beta (\gamma)} \left[ J\vec{S}_i\cdot\vec{S}_j + K S_i^\gamma S_j^\gamma + \Gamma \left(S_i^\alpha S_j^\beta + S_i^\beta S_j^\alpha \right) \right],
\end{eqnarray*}
where $\gamma$ denotes the bond direction ($x, y$ or $z$) and $\alpha, \beta$ denote the remaining two directions. This model was first studied on the two dimensional honeycomb lattice, using a combination of exact diagonalisation and classical analysis. In the classical analysis, two magnetically ordered phases appear which are not present in the HK model. Firstly, a coplanar spiral phase with ordering wavevector $q=K$ is found, in which the spins lie perpendicular to the [111] direction. Secondly, an incommensurate coplanar spiral phase is also present, with a wavevector which varies throughout the phase. In the exact diagonalisation, very similar phases emerge, with slightly shifted phase boundaries.

\begin{figure}[h!]
  \centering
  \setlength\fboxsep{0pt}
  \setlength\fboxrule{0pt}
  \subfloat[][Phase diagram for the \harzero{} lattice.]{
    \label{fig:h0_spiral}
    \fbox{\includegraphics[scale=.95]{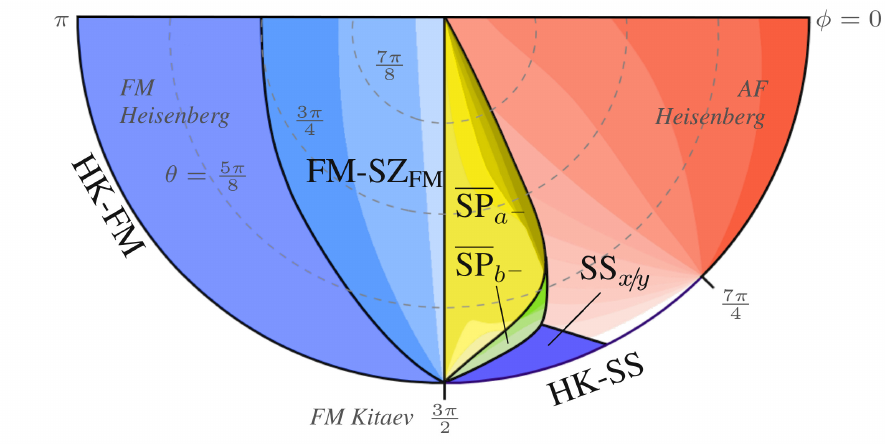}}
  }

  \subfloat[][Phase diagram for the \har{} lattice.]{
    \label{fig:h1_spiral}
    \fbox{\includegraphics[scale=.95]{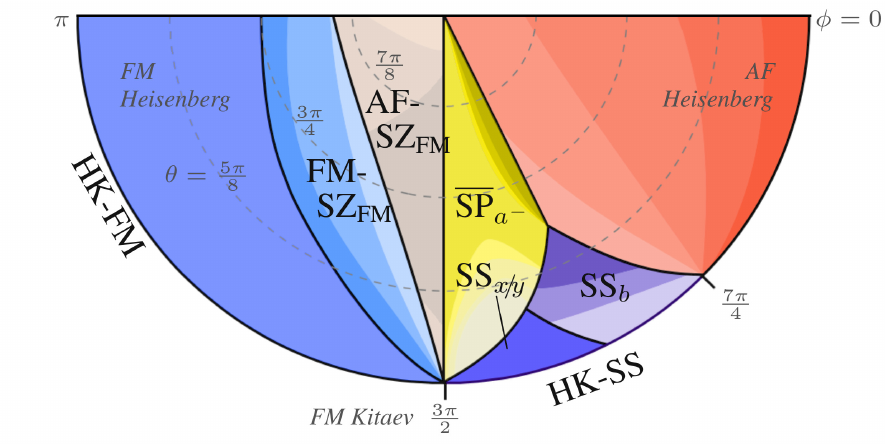}}
  }
  \caption{\label{fig:latticesorder}
  {\bf The classical phase diagrams of the (a) \harzero{} and (b) \har{} latttices.} We use the parameterisation $J=\cos{\phi}\sin{\theta},$ $k=\sin{\phi}\sin{\theta},$ $\Gamma = \cos{\theta}$. In both cases, we restrict to the experimentally relevant potion of the phase diagram, in which K<0 and $\Gamma$<0. In this parameter regime, we find the $\overline{SP}_a$ phase in both models, coloured yellow in the phase diagram. The full phase diagram is discussed in Ref \onlinecite{lee2015theory}.
}
\end{figure}

The local environment of the iridium atoms in the three dimensional crystals \bliiro{} and \gliiro{} resemble those of \nairo{} and \aliiro. As such, a similar symmetric off diagonal exchange term appears in three dimensions, with a factor of $\pm 1$ on different bonds.\cite{lee2015theory} Being a three-dimensional model, numerical methods for analysing the full quantum spin model are infeasible. However, the classical spin model can still be analysed on the \harzero{} and \har{} lattices, and the classical magnetic orders identified. This is an important starting point for the analysis of these systems, and results from the 2-d honeycomb structure indicate the strong possibility that these results are highly relevant to the quantum model as well.

The addition of the $\Gamma$ term to the HK model has significant effects on the possible magnetic orders. Each of the magnetic phases present in the HK model extend to finite $\Gamma$, with some showing continuous deformation of the spin directions throughout the phase. A number of new phases are present as well, including four non-coplanar spirals states, two multiple-Q states, a new antiferromagnetic state and a state with finite net moment. Each of these is present on both the \harzero{} and \har{} lattices, with the exception of one spiral phase, which is only found on the \harzero{} lattice. In addition, a finite region in which the Kitaev spin liquid phase is stable must be present, as the $\Gamma$ term also consists of two spin interactions.

Of these magnetic orders, the non-coplanar spiral states are of the greatest interest, due to their possible relevance to the experimental results. On the \harzero{} lattice one of the ground state magnetic orders, called the $\overline{SP}_a$ phase, has the same symmetries as the experimentally determined magnetic order, and can be shown to be in the same magnetic phase. In addition, the ordering wavevector found experimentally is within the range of those allowed in this phase. The $\overline{SP}_a$ phase is found to be the ground state when the Kitaev term is ferromagnetic, the Heisenberg term is antiferromagnetic and smaller in magnitude than the Kitaev term and the $\Gamma$ term has a particular sign, all of which is consistent with the expectations from the strong coupling expansion. Compounded with this, the $ab-initio$ for \bliiro{} suggest that the parameters have a strong possibility of lying within this region.\cite{lee2015theory,kim2015predominance}

On the \har{} lattice, the ordering wavevector found experimentally is once again found in the range of those allowed in the $\overline{SP}_a$ phase, and this phase is again found in the region predicted by the strong coupling theory. However, the theoretically predicted magnetic order differs from that found in experiment for one component of the spins.

In addition to the HK$\Gamma$ model described above, a model which considers one-dimensional zigzag chains of spins coupled to one another in a three dimensional system has been explored as a possible route through which the magnetic order found in these systems may arise.\cite{kimchi2015unified} This model treats the bonds along two directions (forming the zigzag chains) using a pure HK model, while including symmetric off-diagonal spin exchange on the connecting bonds. A particular version of soft-spin analysis of this model, involving mixing different eigenstates of the classical spin Hamiltonian, is able to reproduce the magnetic order found in experiments on both \bliiro{} and \gliiro{}.

Although the coupled zigzag chain model appears to yield results which are consistent with experiments, it is important to note that the methods described to analyse these models are very different.\cite{lee2015two} It is found that, if the same eigen-mode mixing method is used, the HK$\Gamma$ model produces results consistent with the magnetic order discovered in both $\beta$ and \gliiro, similarly to the coupled zigzag chain model. In addition, a classical analysis of the coupled zigzag chain model finds a spiral phase with a commensurate wavevector, inconsistent with experimental results. A study of the one dimensional quantum limit of this model, in which only the zigzag chains are present, indicates that the $\Gamma$ term within the chains, which is absent in the original zig-zag chain model, is critical for the incommensurate spiral ordering.

\subsection{\label{sec:3d_kitaev}Discussion}

While some details are different in the two models and the two approaches, the similarity of the magnetic phases which emerge from these approaches to experimental results are extremely promising. Additional terms, such as further neighbour interactions, higher spin interaction terms, anisotropies between the different bond directions and further off-diagonal spin interactions could all contribute to some extent, and could serve to improve the agreement with experiment. In addition, quantum effects may be critical to the magnetic order; further studies of these possibilities provide a worthwhile future direction.

One major motivation for the study of the \bliiro{} and \gliiro{} crystals remains the possibility of finding spin liquid physics. In the HK$\Gamma$ model, the spiral phase which is found experimentally in \bliiro{} connects to the ferromagnetic Kitaev spin liquid. Above the ordering temperature, it is possible that signatures of the spin liquid remain, which could be observed in neutron scattering measurements. Further, the $ab-initio$ results indicate that this phase lies close to the spin liquid phase boundary, meaning that a small change in the relative strength of the different interactions may be enough to give rise to a spin liquid phase.\cite{kim2015predominance} A recent study on \bliiro{} has shown that under moderate pressure, evidence of the magnetic order disappears.\cite{takayama2015hyperhoneycomb} The possibility that pressure forces this system into a spin-liquid phase offers another fascinating direction for future research.

\section{Discussion and outlook}

In this review, we focused on recent developments in theoretical and experimental
studies of pyrochlore iridates and 3D honeycomb iridates.
In the case of pyrochlore iridates, we highlighted recent efforts to uncover
novel quantum critical phases and quantum criticality as well as
possible topological phases in new platforms of thin film, surfaces, and domain walls.
We discussed the discovery of 3D honeycomb iridates and theoretical
efforts to understand possible quantum spin liquid phases and
novel spiral magnetic orders identified in experiments.

There has been much notable progress in the studies of other iridates
and 5d transition metal oxides, which have not been discussed so far in this review article.
In particular, a recent ARPES experiment
on Sr$_2$IrO$_4$ with K-doping at the surface reported the presence of
a d-wave-like excitation gap, which could be a signature of the much waited
d-wave superconductivity (SC) in electron-doped Sr$_2$IrO$_4$, which
is an antiferromagnetic insulator in the absence of doping.\cite{kim2015observation}
Previous theoretical studies suggest the emergence of d-wave SC with
electron doping in Sr$_2$IrO$_4$,\cite{Wang2011sc} in analogy to the hole-doped
cuprate La$_2$CuO$_4$ which has the same Fermi surface topology
as the electron-doped $J_{\rm eff}=1/2$ band in Sr$_2$IrO$_4$.
It would also be important to understand why
several other experimental attempts to achieve the electron doping
via cationic substitution have not lead to SC.\cite{Calder2015,yuan2015jeff,ye2015structure} Another interesting avenue
is hole-doping in Sr$_2$IrO$_4$.\cite{Meng2014sc} It has been suggested
that the multi-orbital interactions are inherently more important in the hole-doped
system due to the fact that the hole would go into $J_{\rm eff}=3/2$ bands.
This would lead to a spin-triplet p-wave SC according to a recent theoretical study.

It has also been suggested that SrIrO$_3$ with the orthorhombic
perovskite structure would be an example of a topological crystalline
nodal semi-metal, where the presence of the nodal line spectrum is
topologically protected by the non-symorphic crystal symmetry.\cite{chen2014topological}
While a recent ARPES experiment on an epitaxially grown SrIrO$_3$ film
may be a good starting point for such an investigation, further
experiments on different aspects of the nodal line semi-metal would
be warranted for full understanding of this novel semi-metal phase.\cite{nie2013arpes,Nie2015t}

While much attention has been paid to Na$_2$IrO$_3$ and
$\alpha$-Li$_2$IrO$_3$ for 2D Kitaev physics, recent experiments on
$\alpha$-RuCl$_3$, where Ru$^{3+}$ ions are sitting on a 2D honeycomb lattice,
strongly suggest that similar physics may be at work for this material, even though
Ru is a 4d element and hence has a smaller spin-orbit coupling.\cite{Sandilands2015,sandilands2015orbital,sears2015magnetic,plumb2014alpha,johnson2015monoclinic,kim2015structural,Fang2015some,Kim2014alpha}
It has been suggested that the correlation effects may drive the
system to be in the strong spin-orbit coupled regime.
Moreover, a recent neutron scattering experiment on
a single crystal of $\alpha$-RuCl$_3$ reported the presence
of a continuum of spin excitations at intermediate energy scales,
while the material shows the zig-zag magnetic order at
low temperature.\cite{banerjee2015proximate,2015Naglertalk} The continuum of excitations are interpreted
as a possible signature of
the excitations in the nearby Kiatev spin liquid phase.
If true, this means the Kitaev spin liquid phase is nearby
and one may be able to drive the system to the
spin liquid regime using a suitably chosen external perturbation.

Acknowledgement: We would like to acknowledge discussions and past collaborations with Subhro Bhattacharjee, William Witczak-Krempa, Jeffrey Rau,
Kyusung Hwang, Heung-Sik Kim, Hosub Jin, Yuan-Ming Lu,
Sungbin Lee, Gang Chen, Zi Yang Meng, Jae-Seung Jeong,
Leon Balents, Cenke Xu, Eun-Gook Moon, 
Arun Paramekanti, Hae-Young Kee, 
Young-June Kim, Patrick Clancy, Stephen Julian, Sungbin Lee, 
George Jackeli, Giniyat Khaliullin, Ashvin Vishwanath, T. Senthil, 
Simon Trebst, Lucile Savary,
Satoru Nakatsuji, Hidenori Takagi, B. J. Kim, Gang Cao, Radu Coldea,
Philipp Gegenwart, Yogesh Singh,
Roser Valenti, Natalia Perkins, Yukitoshi Motome,
Masatoshi Imada, Naoto Nagaosa, Roderich Moessner,
Stephen Nagler, Greg Fiete, Michel Gingras, Bruce Gaulin, Yoshinori Tokura and Masashi Kawasaki.
This work was supported
by the NSERC of Canada, the Canadian Institute
for Advanced Research, and the Center for Quantum Materials
at the University of Toronto. BJY was supported by IBS-R009-D1 and Research
Resettlement Fund for the new faculty of Seoul National University.
YBK would like to thank the Kavli Institute for Theoretical Physics where
part of this work was done. The work at KITP was supported in part by NSF
Grant No. NSF PHY11-25915.

\bibliography{review}

\end{document}